\documentclass[11pt,a4paper]{article}
\pdfoutput=1

\usepackage[utf8]{inputenc}
\usepackage[T1]{fontenc}
\usepackage{lmodern}
\usepackage[margin=1in]{geometry}

\usepackage{amsmath,amssymb,amsthm}
\usepackage{mathtools}
\usepackage{xspace}
\usepackage{graphicx}
\usepackage{algorithm}
\usepackage[noend]{algpseudocode}
\usepackage{threeparttable}
\usepackage{tabularx}
\usepackage{enumerate}
\usepackage{thm-restate}
\usepackage[unicode,pdfencoding=auto]{hyperref}
\usepackage{bookmark}

\graphicspath{{./}}
\theoremstyle{plain}
\newtheorem{theorem}{Theorem}[section]
\newtheorem{lemma}[theorem]{Lemma}
\newtheorem{corollary}[theorem]{Corollary}

\theoremstyle{definition}
\newtheorem{definition}[theorem]{Definition}
\newtheorem{remark}[theorem]{Remark}
\makeatletter
\AtBeginDocument{\@for\thm@env:=theorem,lemma,corollary,proposition,definition,remark\do{%
    \expandafter\def\csname theH\thm@env\endcsname{\thesection.\arabic{theorem}}}}
\makeatother

\allowdisplaybreaks[2]
\newcommand{\HYBRIDz}{\textsf{HYBRID}\ensuremath{_0}\xspace}
\newcommand{\CONGEST}{\textsf{CONGEST}\xspace}
\newcommand{\NCCz}{\textsf{NCC}\ensuremath{_0}\xspace}
\newcommand{\ID}{\mathrm{ID}}

\title{Improved (Minimum) Spanning Tree Construction in the \textsf{HYBRID} Model}
\renewcommand{\and}{\end{tabular}\hskip 1.5em\begin{tabular}[t]{c}}
\author{Shoma Hiraoka\\\normalsize The University of Osaka\\\normalsize s-hiraok@ist.osaka-u.ac.jp
    \and Naoki Kitamura\\\normalsize The University of Osaka\\\normalsize n-kitamura@ist.osaka-u.ac.jp
    \and Taisuke Izumi\\\normalsize The University of Osaka\\\normalsize t-izumi@ist.osaka-u.ac.jp}
\date{}

\begin{document}

\maketitle

\begin{abstract}
    In the \HYBRIDz model, vertices exchange $O(\log n)$-bit messages in synchronous rounds, each addressed to a single recipient, both along input edges (\emph{local communication}) and directly to vertices whose identifiers they know (\emph{global communication}).
    The \emph{local capacity} is the maximum number of messages a vertex may send per round to each input neighbor by local communication, and the \emph{global capacity} is the maximum number of messages a vertex may send, and also the maximum number it may receive, per round by global communication.
    The standard setting, which we call the \emph{standard capacity}, is local capacity $O(1)$ and global capacity $O(\log n)$.
    A well-formed tree (\textsf{WFT}) is a virtual tree of maximum degree $O(1)$ and diameter $O(\log n)$.
    Known algorithms construct a spanning tree (\textsf{ST}) and a \textsf{WFT} either in $O(\log n)$ rounds with global capacity $O(\log^2 n)$ or with standard capacity in $O(\log^2 n)$ rounds.
    We give a randomized algorithm that constructs an \textsf{ST} and a \textsf{WFT} in every connected component of an arbitrary graph in $O(\log n)$ rounds with high probability under standard capacity.

    Our approach is based on the notion of a \emph{$\Delta$-pseudo-spanning representation} ($\Delta$-PSR): a virtual network of maximum degree $\Delta$ together with restoration information, such that any spanning tree of the virtual network can be restored to a spanning tree of the input graph in $O(\log n)$ rounds.
    Trees on such a network can be constructed in $O(\log n)$ rounds with global capacity $O(\Delta\log n)$.
    The $O(\log n)$-round algorithm of G\"{o}tte et al.\ can be viewed as constructing an $O(\log n)$-PSR, which accounts for its global capacity $O(\log^2 n)$.
    Our main technical contribution is to construct an $O(1)$-PSR in $O(\log n)$ rounds with standard capacity, by grouping vertices into clusters of size $\Theta(\log n)$ whose members share responsibility for edges, while an input-edge forest preserves the connectivity needed for restoration.

    As applications, for an input with $k$ distinct edge weights and any integer $1\le p\le k$, we construct an exact minimum spanning tree in $O((k/p)\log n)$ rounds with local capacity $O(p)$ and global capacity $O(p\log n)$, and the same tradeoff applies to $(1+\epsilon)$-approximate minimum spanning trees.
    We also compute biconnected components, articulation points, and bridges in $O(\log n)$ rounds with standard capacity, reducing the global capacity of the previous $O(\log n)$-round algorithm from $O(\log^5 n)$.
\end{abstract}


\section{Introduction}

\subsection{Background and motivation}
\label{subsec:background}

Many real networks are equipped with high-bandwidth direct links to nearby vertices and a low-bandwidth shared infrastructure that reaches distant vertices.
In distributed computing theory, this combination is formalized as the \HYBRIDz model, which combines these two communication layers in a single synchronous model~\cite{augustine2020shortest,feldmann2021fast}.
The input is an undirected graph $G=(V,E)$ on $n$ vertices with an edge-weight function $w$.
Every vertex has a unique identifier (ID), and the two endpoints of each input edge initially know each other's ID and the weight of that edge.
In each round, every vertex can send messages along input edges by \emph{local communication} and directly to any vertex whose ID it knows at the beginning of the round by \emph{global communication}.
Two vertices that know each other's IDs may communicate over a \emph{virtual edge} realized by global communication, regardless of whether they share an input edge.
Global communication allows global tasks to make progress without depending on the diameter of the input graph, but the number of messages a vertex can handle in one round is limited.
Each message carries $O(\log n)$ bits and has a single recipient.
Neither mode provides a broadcast primitive.
The \emph{local capacity} $\lambda$ is the number of messages that a vertex may send per round to each of its input neighbors.
As in \CONGEST, the messages sent to different neighbors may differ.
The \emph{global capacity} $\gamma$ bounds, per round, both the number of global messages a vertex sends and the number of global messages addressed to it.
Every message counts separately even if several messages have the same content.
Thus, sending the same content to two vertices counts as two sent messages, and receiving the same content from two vertices counts as two received messages.\footnote{In particular, global communication is not a (broadcast) congested clique: a vertex can address only vertices whose IDs it knows, and it can send at most $\gamma$ messages per round in total, not one message to each vertex.}
We write $(\lambda,\gamma)$-\HYBRIDz for the model with local capacity $\lambda$ and global capacity $\gamma$.
The original setting of the \HYBRIDz model introduced by Augustine et al.\ assumes $\gamma = O(\log n)$.
We call it the \emph{standard global capacity} in what follows~\cite{augustine2020shortest,feldmann2021fast}.
We also call the combination of local capacity $O(1)$ and the standard global capacity, i.e., $(1,\log n)$-\HYBRIDz, the \emph{standard capacity}.

This paper studies the complexity of constructing tree structures in the \HYBRIDz model.
The problem has several formulations.
The most standard formulation is to construct a spanning tree of the input graph (\textsf{ST}).
Another formulation is to construct a low-diameter spanning tree using virtual edges realized by global communication.
In particular, a tree of maximum degree $O(1)$ and diameter $O(\log n)$ is called a \emph{well-formed tree} (\textsf{WFT}).
For edge-weighted input graphs, minimum spanning tree (\textsf{MST}) construction is also studied.

Previous work shows that a \textsf{WFT} can be constructed in $\operatorname{polylog}(n)$ rounds with the standard global capacity~\cite{chang2025overlay,gmyr2017distributed}.
Combining Bor\r{u}vka-style merging with this gives \textsf{ST} and \textsf{MST} algorithms in $\operatorname{polylog}(n)$ rounds as well~\cite{feldmann2021fast,gmyr2017distributed}.
On the other hand, \textsf{WFT} construction requires $\Omega(\log n)$ rounds however large the capacities are: messages go only to known IDs, so after $t$ rounds vertices that know each other are at input distance at most $2^t$, whereas a \textsf{WFT} of diameter $O(\log n)$ on an input path has an edge spanning distance $\Omega(n/\log n)$.
Angluin et al.\ prove a bound of this type for sorted-list construction~\cite{angluin2005fast}.
Hence the primary question is to optimize the rounds and capacity within the $\operatorname{polylog}(n)$ range.
The best known round complexity for \textsf{ST} and \textsf{WFT} construction is achieved by the algorithms of G\"{o}tte et al.~\cite{gotte2023time} (combined with the degree reduction of Chang et al.~\cite{chang2025overlay} for \textsf{WFT} construction), which run in $O(\log n)$ rounds with high probability.\footnote{Throughout this paper, ``with high probability'' (\emph{w.h.p.}) means with probability at least $1-n^{-\Omega(1)}$, where the hidden constant can be made arbitrarily large.}
However, their construction requires global capacity $O(\log^2 n)$.
Chang et al.\ construct a \textsf{WFT} with the standard global capacity, but their algorithm requires $O(\log^2 n)$ rounds~\cite{chang2025overlay}.
That is, no known construction simultaneously achieves $O(\log n)$ rounds and the standard global capacity.
As for \textsf{MST} construction, Feldmann et al.\ give a deterministic exact algorithm that runs in $O(\log^2 n)$ rounds on \HYBRIDz with standard capacity~\cite[Lemma~19]{feldmann2021fast}.\footnote{The cited paper presents this algorithm in a section on sparse graphs, and obtains it by modifying the overlay construction of Gmyr et al.~\cite[Observation~4]{gmyr2017distributed}, which is stated for input graphs of maximum degree $\operatorname{polylog}(n)$. Neither restriction is essential in $(1,\log n)$-\HYBRIDz. The degree assumption is a model assumption of~\cite{gmyr2017distributed}, where local and global communication share a single $\operatorname{polylog}(n)$ budget per vertex. In \HYBRIDz, the steps of the overlay construction whose cost grows with the degree, namely announcing the current supernode ID to all neighbors, sending merge requests, and arranging the senders of merge requests into a child-sibling path, send $O(1)$ messages per input edge and are thus covered by the local capacity, while every other step runs on overlay trees of constant degree or on a pointer-jumping ring in which each vertex acquires $O(\log n)$ chords. The sparsity of the input is not used in the analysis. The $O(\log^2 n)$ bound therefore holds for input graphs of arbitrary degree.}
Gmyr et al.\ instead construct an exact \textsf{MST} via the subgraphs of edges up to each weight threshold (threshold subgraphs) in $O(W\log^2 n)$ rounds for integer weights at most $W$~\cite{gmyr2017distributed}.
The threshold subgraph changes only at the $k$ distinct edge weights, which makes $k$ a natural parameter.
We ask whether additional capacity allows several thresholds to be processed in parallel.
Table~\ref{tab:related} summarizes the main known results and the results of this paper, together with the model of each reference.
Comparisons with results in related models are discussed in \S\ref{subsec:related}.

\begin{table}[t]
    \centering
    \caption{Main known results and our results on \textsf{WFT}, \textsf{ST}, and \textsf{MST} construction.
        exp. denotes expected time.
        Other randomized bounds hold with high probability.
        Here $d$ is the maximum input degree, $W$ is the maximum weight, and $k$ is the number of distinct edge weights.
        The variable $p$ is the design parameter controlling the parallelism.
        Our exact \textsf{MST} result assumes $1\le p\le k$.
        Our $(1+\epsilon)$-approximate \textsf{MST} result assumes $0<\epsilon\le1$ and $1\le p\le\epsilon^{-1}\log W$.
        Models are restated in $(\lambda,\gamma)$-\HYBRIDz notation.
        Gmyr et al.\ assume $d=\operatorname{polylog}(n)$, and their capacities are our estimates.}
    \label{tab:related}
    \small
    \setlength{\tabcolsep}{4pt}
    \begin{tabularx}{\textwidth}{@{}llcl>{\raggedright\arraybackslash}X@{}}
        \hline
        Reference                                              & Output                               & Det/Rand & Rounds                             & $(\lambda,\gamma)$ \\
        \hline
        Angluin et al.~\cite{angluin2005fast}                  & \textsf{WFT}                         & Rand.    & $O((d+\log n)\log n)$ (exp.)       & $(0,d)$            \\
        Gmyr et al.~\cite{gmyr2017distributed}                 & \textsf{WFT}                         & Det.     & $O(\log^2 n)$                      & $(1,\log n)$       \\
        G\"{o}tte et al.~\cite{gotte2023time,chang2025overlay} & \textsf{WFT}, \textsf{ST}            & Rand.    & $O(\log n)$                        & $(1,\log^2 n)$     \\
        Chang et al.~\cite{chang2025overlay}                   & \textsf{WFT}                         & Rand.    & $O(\log^2 n)$                      & $(1,\log n)$       \\
        \textbf{Our result}                                    & \textsf{WFT}, \textsf{ST}            & Rand.    & $O(\log n)$                        & $(1,\log n)$       \\
        \hline
        Gmyr et al.~\cite{gmyr2017distributed}                 & \textsf{MST}                         & Det.     & $O(W\log^2 n)$                     & $(1,\log n)$       \\
                                                               & $(1+\epsilon)$-approx.\ \textsf{MST} & Det.     & $O(\epsilon^{-1}\log W\log^2 n)$   & $(1,\log n)$       \\
        Feldmann et al.~\cite{feldmann2021fast}                & \textsf{MST}                         & Det.     & $O(\log^2 n)$                      & $(1,\log n)$       \\
        \textbf{Our result}                                    & \textsf{MST}                         & Rand.    & $O((k/p)\log n)$                   & $(p,p\log n)$      \\
                                                               & $(1+\epsilon)$-approx.\ \textsf{MST} & Rand.    & $O((\epsilon^{-1}\log W/p)\log n)$ & $(p,p\log n)$      \\
        \hline
    \end{tabularx}
\end{table}

\subsection{Main results}
\label{subsec:main-results}
Based on the preceding discussion, we focus on the following two questions.
\begin{enumerate}
    \item \label{q:st}
          For input graphs of arbitrary degree, can an \textsf{ST} of each connected component be constructed in $O(\log n)$ rounds with the standard global capacity?
    \item \label{q:mst}
          Can exact and $(1+\epsilon)$-approximate \textsf{MST}s be constructed in $o(\log^2 n)$ rounds by exploiting the number of distinct edge weights and the communication capacity?
\end{enumerate}

Our main contribution is to answer both questions affirmatively under the conditions stated below.
For Question~\ref{q:st}, we give the following theorem.

\begin{theorem}
    \label{thm:low-capacity-st}
    In $(1,\log n)$-\HYBRIDz, an \textsf{ST} and a \textsf{WFT} can be constructed in every connected component of an arbitrary-degree input graph in $O(\log n)$ rounds with high probability.
\end{theorem}
The theorem also applies to any spanning subgraph, i.e., to any subset of the input edges known to both endpoints: run the algorithm using only those edges, treat a vertex incident to none of them as isolated, and keep the original value of $n$.
This restriction does not increase the local or global communication at any vertex.

For Question~\ref{q:mst}, we obtain a tradeoff between capacity and rounds that depends on the number $k$ of distinct edge weights.
The tradeoff is controlled by the number $p$ of distinct weights processed simultaneously, which we call the \emph{parallelism}.
We formulate the result as a reduction that uses only an \textsf{ST} construction algorithm as a black box.

\begin{theorem}
    \label{thm:kweight-mst}
    Let $\gamma=\Omega(\log n)$.
    Suppose that an algorithm constructs a rooted \textsf{ST} consisting only of input edges in each connected component of any subgraph of the input graph in $O(\log n)$ rounds with high probability in $(\lambda,\gamma)$-\HYBRIDz.
    Then, for a connected input graph with $k$ distinct edge weights and any integer $p$ with $1\le p\le k$, an \textsf{MST} can be constructed in $O((k/p)\log n)$ rounds with high probability in $(p(\lambda+1),p\gamma)$-\HYBRIDz.
\end{theorem}

Combining this reduction with Theorem~\ref{thm:low-capacity-st} gives the following result.

\begin{corollary}
    \label{cor:kweight-mst}
    Let $G$ be connected with $k$ distinct edge weights.
    For any integer $1\le p\le k$, an \textsf{MST} can be constructed in $O((k/p)\log n)$ rounds w.h.p.\ in $(p,p\log n)$-\HYBRIDz.
\end{corollary}

Thus, multiplying both capacities by $p$ divides the round bound by $p$: standard capacity ($p=1$) gives $O(k\log n)$ rounds, improving the known $O(\log^2 n)$ bound when $k=o(\log n)$, while $p=k$ gives $O(\log n)$ rounds.

Rounding positive edge weights to powers of $(1+\epsilon)$, while leaving weight $0$ unchanged, extends the same tradeoff to $(1+\epsilon)$-approximate \textsf{MST}: the number of distinct rounded weights replaces $k$ in Corollary~\ref{cor:kweight-mst}.

\begin{corollary}
    \label{cor:approx-mst}
    Let $G=(V,E)$ be connected, and let $W\coloneqq\max(\{2\}\cup\{w(e):e\in E\})$.
    Suppose that $0<\epsilon\le1$ is given to all vertices and that each vertex can compute and compare rounded weights exactly by local computation.
    For any integer $p$ with $1\le p\le\epsilon^{-1}\log W$, a $(1+\epsilon)$-approximate \textsf{MST} can be constructed in $O((\epsilon^{-1}\log W/p)\log n)$ rounds w.h.p.\ in $(p,p\log n)$-\HYBRIDz.
\end{corollary}

As a further application, we combine our \textsf{ST} and \textsf{WFT} constructions with the Tarjan--Vishkin reduction to compute biconnected components (BCCs), articulation points, and bridges with standard capacity.
The previous $O(\log n)$-round implementation of G\"{o}tte et al.\ requires global capacity $O(\log^5 n)$~\cite{gotte2023time}.

\begin{theorem}
    \label{thm:biconnectivity}
    On any undirected input graph, the following can be computed in $O(\log n)$ rounds with high probability in $(1,\log n)$-\HYBRIDz.
    Both endpoints of each input edge learn the label of the BCC containing that edge and whether the edge is a bridge.
    Each vertex learns whether it is an articulation point.
\end{theorem}

\subsection{Technical overview}
\label{subsec:overview}

\subsubsection{Overall approach}
\label{subsubsec:outline-representation}
Our construction builds on the tree construction of G\"{o}tte et al.\ for arbitrary-degree graphs~\cite{gotte2023time}.
We call the graph obtained by treating global communication as virtual edges a \emph{virtual network}.
For \textsf{ST} construction, their algorithm consists of the following three steps.
\begin{enumerate}
    \item Construct a virtual network that spans the input graph and has maximum degree $O(\log n)$.
    \item Construct an \textsf{ST} of this network by an algorithm for graphs of maximum degree $\Delta$ that runs in $O(\log n)$ rounds in $(1,\Delta\log n)$-\HYBRIDz (Theorem~\ref{thm:known-tree-construction}).
    \item The spanning tree from Step~2 may contain virtual edges, so convert it into a spanning tree consisting only of input edges.
\end{enumerate}
Since the virtual network from Step~1 has maximum degree $O(\log n)$, Step~2 runs with $\Delta=O(\log n)$ and requires global capacity $O(\log^2 n)$.
This is the bottleneck in their construction, and reducing the maximum degree in Step~1 directly reduces the capacity.

To abstract Steps~1 and~3, we introduce the notion of a \emph{$\Delta$-pseudo-spanning representation} ($\Delta$-PSR).
Informally, a $\Delta$-PSR of a connected input graph $G$ consists of a virtual network $H$ spanning $G$ with maximum degree at most $\Delta$, side information stored at each vertex, and a restoration algorithm that takes an arbitrary spanning tree of $H$ together with the side information and returns a spanning tree consisting only of edges of $G$.
The restoration algorithm must run in $O(\log n)$ rounds w.h.p.\ in $(1,\Delta\log n)$-\HYBRIDz.
Thus, a $\Delta$-PSR specifies the structure to be produced in Step~1 together with the requirement that the conversion in Step~3 be possible.
The construction of G\"{o}tte et al.\ can be interpreted as constructing an $O(\log n)$-PSR with standard capacity.

Our technical focus is therefore to construct an $O(1)$-PSR with standard capacity in $O(\log n)$ rounds w.h.p.\ (\S\ref{subsubsec:outline-st}).
Applying Steps~2 and~3 to it gives Theorem~\ref{thm:low-capacity-st}.

\subsubsection{Constructing an \texorpdfstring{$O(1)$}{O(1)}-pseudo-spanning representation}
\label{subsubsec:outline-st}
Our construction of an $O(1)$-PSR extends the construction of an $O(\log n)$-PSR by G\"{o}tte et al.~\cite[Theorems~1.2 and~1.4]{gotte2023time}.
We first describe the idea behind their construction.
In the first stage, sparsification based on the randomized clustering of Miller--Peng--Xu, which we call MPX-based sparsification~\cite{miller2013parallel}, produces a spanning subgraph preserving the connectivity of the input graph and an orientation of its edges.
Under this orientation, every vertex has outdegree $O(\log n)$ with high probability~\cite[Lemma~4.2]{gotte2023time}.
We use only these two output properties and treat the sparsification as a black box.
The indegree is not bounded: $\omega(\log n)$ edges may be directed toward the same vertex.

\begin{figure}[htbp]
    \centering
    \includegraphics[width=1.00\textwidth]{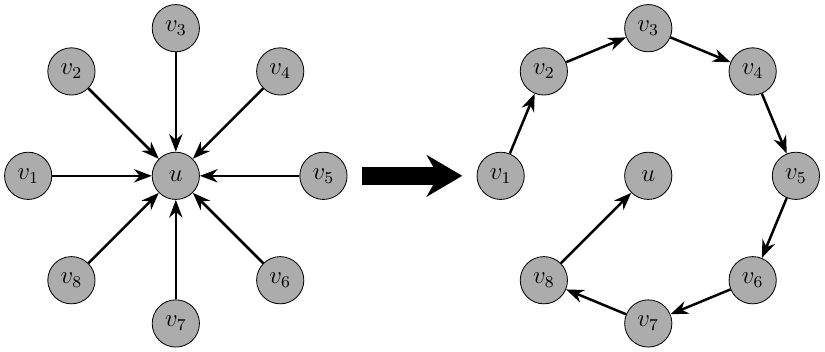}
    \caption{Rewiring incoming edges. A path connects the incoming neighbors, and only one original incoming edge to the central vertex is retained.}
    \label{fig:gotte-degree-reduction}
\end{figure}

In the second stage, virtual edges reduce the number of original edges entering each vertex $u$ of indegree $\omega(\log n)$ to at most one.
Specifically, the tails of the edges entering $u$, its incoming neighbors, are connected by a path of virtual edges.
Only one of these vertices remains connected to $u$, and the other edges entering $u$ are removed from the virtual network (Figure~\ref{fig:gotte-degree-reduction}).
Every vertex incident to a new virtual edge in this transformation is an incoming neighbor of $u$ and shares an input edge with $u$.
Therefore, $u$ can use local communication on its input edges both to exchange the IDs needed to realize the virtual edges and to restore the removed input edges that are needed during $\Delta$-PSR restoration.
Both operations take $O(1)$ rounds.
Each original outgoing edge of a vertex $v$ is replaced by at most two edges, and the indegree of $v$ increases by at most one for each of its original outgoing edges.
Thus, the resulting virtual network has both indegree and outdegree $O(\log n)$, giving an $O(\log n)$-PSR.

Our task is to reduce the degree of this virtual network from $O(\log n)$ to $O(1)$.
The construction has three main ingredients.
\begin{itemize}
    \item \emph{Clusters.}
          We grow an input-edge forest $F$ whose trees have at least $\lceil\log n\rceil$ vertices and cut each tree bottom-up into vertex sets of size $\Theta(\log n)$, called \emph{clusters} (\S\ref{subsubsec:forest-and-clusters}).
    \item \emph{Sparsification between clusters.}
          We run MPX-based sparsification on the graph obtained by contracting each cluster to a single vertex, where two clusters are adjacent if an input edge joins them.
          This contracted graph is not a communication network, and each step of the sparsification is simulated over the input edges between members.
          Afterwards, each cluster keeps $O(\log n)$ input edges to other clusters, called \emph{representative input edges}.
    \item \emph{Load balancing.}
          For each representative input edge, one of the two clusters it joins is designated as \emph{responsible} for it.
          Each cluster distributes the $O(\log n)$ edges it is responsible for evenly among its $\Theta(\log n)$ members.
          The member that receives an edge $e$ is the \emph{responsible vertex} of $e$ and represents $e$ in the virtual network in place of the endpoints of $e$.
          Finally, a \textsf{WFT}~\cite[Appendix~D.1]{chang2025overlay} on each tree of $F$ lets the members of a cluster communicate over constant-degree virtual edges.
\end{itemize}
Each vertex is thus responsible for $O(1)$ edges, giving a virtual network of constant degree.

We next describe the restoration algorithm, which converts a spanning tree of the $O(1)$-PSR into a spanning tree consisting only of input edges.
Two issues need to be addressed.
\begin{itemize}
    \item A cluster need not induce a connected subgraph of the input graph.
          A partition into clusters of size $\Theta(\log n)$, each connected by input edges within the cluster, need not exist: in a star, for example, two or more leaves without the center cannot be connected using input edges within that set.
    \item Connecting the vertices of a cluster may therefore require input edges outside the cluster, which may create cycles unless they are selected carefully.
\end{itemize}
Since each cluster lies within one tree of $F$, whose input edges already connect its vertices, restoration only has to connect the trees of $F$ by representative input edges without creating cycles.
For this purpose we introduce \emph{quotient-tree selection} (Lemma~\ref{lem:quotient-tree-selection}).
Given a rooted tree and a partition of its vertices, it takes, in each set except the one containing the root, the vertex of minimum depth and selects the edge from it to its parent.
The parent is shallower, so it lies in another set whose minimum depth is smaller.
Following selected edges therefore strictly decreases the minimum depth and reaches the set containing the root without cycles, so the selected edges form a spanning tree of the quotient graph obtained by contracting each set.
For restoration, we first transform the spanning tree of the virtual network so that its edges between different trees of $F$ correspond one-to-one to representative input edges (\S\ref{subsec:arbitrary-degree-tree}), and then apply quotient-tree selection with the sets corresponding to the trees of $F$.
Replacing the selected edges by their representative input edges and adding $F$ gives an input-edge spanning tree.
The rule is not specific to $F$, and we reuse it for \textsf{MST} below.

\subsubsection{Generalization to few-weight \textsf{MST}}
\label{subsubsec:outline-reduction}
Our construction is based on the Kruskal-style \textsf{MST} construction of Gmyr et al.\ using threshold subgraphs~\cite{gmyr2017distributed}.
Let $w_1<\cdots<w_k$ be the distinct edge weights of a connected input graph $G=(V,E)$.
Let $G_i$ be the subgraph consisting of edges of weight at most $w_i$, and let $G_0\coloneqq(V,\emptyset)$.
At weight $w_i$, it suffices to select edges that connect the $G_{i-1}$-components (the connected components of $G_{i-1}$) within each $G_i$-component into a tree.

This is again quotient-tree selection: we construct a rooted \textsf{ST} $T$ in each $G_i$-component and, in each contained $G_{i-1}$-component except the one with the root, select the edge of $T$ from its minimum-depth vertex to that vertex's parent.
Each selected edge lies in $G_i$ and joins distinct $G_{i-1}$-components, so it has weight $w_i$, and the union of the selected edges over all weights is an \textsf{MST}.
The \textsf{WFT}s used for aggregation are constructed from the resulting \textsf{ST}s.

With parallelism $p$, we multiply both capacities by $p$.
The processing of each weight is independent of the edges selected at other weights, so processing at most $p$ distinct weights in parallel reduces the number of rounds.
We prove correctness and analyze communication and rounds, including the dissemination of weights, in \S\ref{subsec:kweight-reduction}.

\subsubsection{Biconnected components}
Tarjan and Vishkin reduce biconnectivity to connectivity~\cite{tarjan1985efficient}.
Given a rooted \textsf{ST} $T$, they define an auxiliary graph $H_{\mathrm{BCC}}$ whose vertices are the edges of $T$, such that two tree edges belong to the same biconnected component if and only if they are connected in $H_{\mathrm{BCC}}$.
Deciding which auxiliary edges exist requires, at every vertex, aggregate values over its subtree, namely the minimum and maximum preorder numbers reachable from the subtree by one non-tree edge.
G\"{o}tte et al.\ implemented this reduction in $O(\log n)$ rounds with global capacity $O(\log^5 n)$~\cite{gotte2023time}.
Using our \textsf{ST} and \textsf{WFT} constructions, we achieve the same round bound with standard capacity.

Their implementation already simulates each auxiliary edge over an input edge, so the auxiliary graph inherits the local communication of the input graph, and we reuse this embedding.
Two of their steps exceed standard capacity: the \textsf{ST} and \textsf{WFT} constructions, and the aggregation of the Tarjan--Vishkin subtree values through multicast trees.
We replace the former by Theorem~\ref{thm:low-capacity-st} and compute the latter by aggregation over an Euler tour in which each vertex holds $O(1)$ items, so every step runs with standard capacity.
Details, including the detection of articulation points and bridges, are given in \S\ref{sec:bcc} and Appendix~\ref{app:bcc}.

\subsection{Further related work}
\label{subsec:related}
To compare results across models, we call the number of local and global messages processed by a vertex per round its \emph{communication capacity}.
For maximum input degree $d$, standard capacity gives communication capacity $O(d+\log n)$.
The \textsf{ST} and \textsf{WFT} constructions of G\"{o}tte et al.\ use global capacity $O(\log^2 n)$, so a vertex of degree $d$ processes $\Theta(d+\log^2 n)$ messages per round. They pose eradicating the additional $\log n$ factor, i.e., attaining $\Theta(d+\log n)$, as an open goal~\cite{gotte2023time}.
Our construction attains it in $O(\log n)$ rounds, whereas Chang et al.\ instead improve the total message complexity of \textsf{WFT} construction~\cite{chang2025overlay}.
We do not bound the total number of messages independently of $|E|$, so achieving $O(\log n)$ rounds and $O(n\operatorname{polylog}n)$ total messages simultaneously remains open.

In the standard \textsf{NCC}, all vertex IDs are known initially.
Augustine et al.\ construct an \textsf{MST} in $O(\log^4 n)$ rounds~\cite{augustine2019distributed}, and Nowicki studies spanning forests and minimum spanning forests in weighted graphs~\cite{nowicki2018random}.
These algorithms do not directly apply to \NCCz, which is equivalent to $(0,\log n)$-\HYBRIDz and in which messages may be addressed only to known IDs, and their initial-knowledge assumptions differ from ours.

Hybrid models have also been studied from a complexity-theoretic perspective.
Fraigniaud et al.\ classify the computing power of models that combine a constant number of rounds of \textsf{LOCAL} or \CONGEST with rounds of the broadcast congested clique~\cite{fraigniaud2022computing}.
In their global mode, each vertex broadcasts to all vertices, whereas global communication in \HYBRIDz is unicast, restricted to known IDs, and limited by the global capacity.

\section{Preliminaries}
\label{sec:preliminaries}
We first introduce the formal definition of $(\lambda,\gamma)$-\HYBRIDz.
The input is a simple undirected graph $G=(V,E)$ with $n=|V|$ and an edge-weight function $w:E\to\{0,\ldots,\operatorname{poly}(n)\}$.
We also write $V(H)$ for the vertex set of a graph $H$.
Each vertex has a unique $O(\log n)$-bit identifier, knows $n$, and initially knows the ID of every input neighbor and the weight of the input edge joining them.
Algorithms proceed in synchronous rounds.
In each round, vertices send and receive messages, and perform unrestricted local computation.
The size of each message is bounded by $O(\log n)$ bits.
As explained in the introduction, there are two communication modes, called \emph{local communication} and \emph{global communication}.
In local communication, each vertex can send up to $\lambda$ messages per round to each input neighbor, and the messages sent to different neighbors are independent.
In global communication, each vertex can send up to $\gamma$ messages per round in total.
Each of these messages is addressed to a single vertex whose ID the sender knows at the beginning of the round.
Several of them may have the same recipient, and messages with identical content sent to different recipients count as different messages.
The number of global messages addressed to a vertex in one round, counted with multiplicity regardless of content, is its \emph{receive count}, and it is bounded by $\gamma$ as well.
The parameters $\lambda$ and $\gamma$ are respectively referred to as \emph{local capacity} and \emph{global capacity}.
We assume that both modes can be used in every round.
If the number of messages a vertex sends, or its receive count, exceeds $\gamma$ in one round, an arbitrary subset of $\gamma$ of the messages concerned is delivered and the rest are discarded.
Every message that is not discarded is guaranteed to arrive at its destination in the same round.
An ID newly received in round $r$ may be used as a destination only in round $r+1$ or later.
Constant factors in $\lambda$ and $\gamma$ are absorbed into asymptotic notation.

Global communication between two vertices that know each other's IDs realizes a virtual edge, and adding virtual edges to the input graph yields a \emph{virtual network}:
\begin{definition}[Virtual network]
    \label{def:virtual-network}
    Let $U\subseteq V$.
    A \emph{virtual network on $U$} is a graph $H=(U,E_H)$, and the edges in $E_H\setminus E$ are called \emph{virtual edges}.
    We say that $H$ is \emph{realized} if, for every virtual edge $\{u,v\}$, vertices $u$ and $v$ know each other's IDs, and every vertex has at most $\gamma$ incident virtual edges.
    A virtual network on $V$ is simply called a virtual network.
\end{definition}
The input graph itself is a realized virtual network with no virtual edges.
On a realized virtual network, messages along input edges use local communication, and each message along a virtual edge is realized by one global message.

A tree is \emph{rooted} if one of its vertices is designated as the \emph{root} and every non-root vertex knows the ID of its parent.
Rooting a tree at a vertex $r$ means choosing $r$ as its root and informing every non-root vertex that its parent is the next vertex on the tree path to $r$.

\begin{definition}[\textsf{WFT} and spanning tree]
    Let $H=(V_H,E_H)$ be a subgraph of the input graph.
    A \emph{well-formed tree} (\textsf{WFT}) on $H$ is a realized virtual network $T$ such that $T$ is a rooted tree with maximum degree $O(1)$ and diameter $O(\log n)$.
    A \emph{spanning tree} (\textsf{ST}) on $H$ is a rooted spanning tree of $H$.
\end{definition}

By slight abuse of terminology, we also use these terms when $H$ is disconnected (meaning a well-formed forest or a spanning forest).

\section{Low-capacity \textsf{WFT} and spanning-tree construction}
\label{sec:low-capacity-tree}
Every statement whose proof is omitted or only sketched below names the appendix subsection that contains its full proof.
Those proofs carry the complete round and capacity accounting.

\subsection{Useful tools}
\label{subsec:constant-degree-known}

Besides virtual edges, our constructions in \S\ref{subsec:arbitrary-degree-tree} and Appendix~\ref{app:known-tree-construction-details} also use auxiliary vertices, called \emph{copies}, each of which is simulated by a vertex of $G$.
The virtual edges incident to a copy are realized by global communication between the simulating vertices.
To distinguish them from copies, we call the vertices of $G$ \emph{real vertices}.
We also use \textsf{WFT} for realized rooted trees on subsets of real vertices, with the same constant-degree and $O(\log n)$-diameter requirements.

\begin{lemma}[{variant of \cite[Appendix~D.1]{chang2025overlay}}]
    \label{lem:tree-to-wft}
    Let $T_U$ be a rooted tree on real vertices $U\subseteq V$ in which the endpoints of every edge know each other's IDs and every vertex has at most $d$ incident virtual edges.
    A \textsf{WFT} on $U$ with the same root can be constructed deterministically in $O(\log n)$ rounds in $(1,d+\log n)$-\HYBRIDz.
    Both endpoints of every output edge know the parent--child relation.
\end{lemma}
\begin{remark}
    \label{rem:tree-to-wft-variants}
    In the construction of Chang et al.~\cite[Appendix~D.1]{chang2025overlay}, only the first step depends on the degree in $T_U$: it sends one message over each edge of $T_U$ and yields a tree of maximum degree three whose adjacent vertices know each other's IDs, on which all later steps use $O(\log n)$ global messages per vertex.
    Hence, if every edge of $T_U$ is virtual, the construction runs in $(0,d+\log n)$-\HYBRIDz.
\end{remark}

\begin{lemma}[{\cite[Lemma~4 and Theorem~5]{feldmann2021fast}}, restated]
    \label{lem:fhs-tree-rooting}
    A realized tree $T_U$ on real vertices $U\subseteq V$ with positive $O(\log n)$-bit integer edge weights known to both endpoints can be rooted at any $r\in U$, each vertex learning its distance from $r$, in $O(\log n)$ rounds: in $(1,\log n)$-\HYBRIDz if $T_U$ has only input edges, and in $(0,\log n)$-\HYBRIDz if its maximum degree is $O(1)$.
\end{lemma}
Setting every edge weight to 1 gives the depth of each vertex.
Next, we formalize quotient-tree selection for a rooted tree $T$ and a partition of $V(T)$ whose sets need not be connected in $T$.

\begin{lemma}[Quotient-tree selection]
    \label{lem:quotient-tree-selection}
    Let $T$ be a tree rooted at $r$ with parent function $p_T$, and let $\mathcal Z$ be a partition of $V(T)$ into nonempty sets.
    For each $Z\in\mathcal Z$, let $z_Z$ be the vertex in $Z$ with the minimum pair of depth and vertex ID, and let $E^\star\coloneqq\{\{z_Z,p_T(z_Z)\}:Z\in\mathcal Z,\ r\notin Z\}$.
    Then $E^\star$ forms a spanning tree of the quotient graph obtained by contracting each set to one vertex, rooted at the set containing $r$.
\end{lemma}
\begin{proof}
    Let $Z\in\mathcal Z$ with $r\notin Z$.
    Then $z_Z\ne r$, and its parent $p_T(z_Z)$ is shallower than $z_Z$, so by the choice of $z_Z$ it lies in another set $Z'$, whose minimum depth is smaller than that of $Z$.
    Hence, following the selected edges from any set strictly decreases the minimum depth, never revisits a set, and reaches the set containing $r$.
    Thus $E^\star$ connects all sets without cycles.
\end{proof}

\begin{theorem}
    \label{thm:known-tree-construction}
    Let $1\le k\le n$, and let $H$ be a realized connected graph on a set $U$ of real vertices with maximum degree at most $k$.
    A rooted spanning tree of $H$ and a \textsf{WFT} on $U$ can be constructed in $(0,k\log n)$-\HYBRIDz in $O(\log n)$ rounds w.h.p.
\end{theorem}
For the spanning tree, each vertex of $H$ creates one copy per incident edge and connects its copies in a path, which gives a graph $\widetilde H$ of maximum degree three in which each vertex simulates at most $k$ copies.
The constant-degree algorithm of Feldmann et al.~\cite[Lemma~12]{feldmann2021fast} constructs a spanning tree of $\widetilde H$ with $O(\log n)$ global messages per copy, hence $O(k\log n)$ per vertex, and quotient-tree selection with one set per path turns it into a spanning tree of $H$ (Appendix~\ref{app:known-tree-construction-details}).
Once the spanning tree is obtained, Lemma~\ref{lem:tree-to-wft} with $d=k$, in the all-virtual variant of Remark~\ref{rem:tree-to-wft-variants}, constructs the \textsf{WFT} with global capacity $O(k+\log n)$, within the stated bound.

\subsection{Pseudo-spanning representation}
\label{subsec:psr-framework}
The next definition states what a virtual network must provide so that Theorem~\ref{thm:known-tree-construction} applies to it and the resulting tree can be converted into a spanning tree of $G$.

\begin{definition}[$\Delta$-pseudo-spanning representation]
    \label{def:pseudo-spanning}
    Let $G=(V,E)$ be a connected graph.
    A \emph{$\Delta$-pseudo-spanning representation} ($\Delta$-PSR) of $G$ is a triple of a realized connected virtual network $H=(V,E_H)$, side information $I:V\to\{0,1\}^{\ast}$, and a restoration algorithm $A$ satisfying the following conditions.
    \begin{enumerate}
        \item The maximum degree of $H$ is at most $\Delta$, counting virtual and input edges.
        \item $A$ takes as input any rooted spanning tree $H'$ of $H$ and $I$, given locally at each vertex, and runs in $(1,\Delta\log n)$-\HYBRIDz in $O(\log n)$ rounds with high probability.
        \item The output is a rooted spanning tree of $G$ in which each vertex knows its incident edges.
    \end{enumerate}
\end{definition}
Since $V(H)=V$, a \textsf{WFT} constructed on $H$ is already a \textsf{WFT} on $V$ and needs no restoration.

\begin{lemma}
    \label{lem:pseudo-spanning-reduction}
    Suppose that a $\Delta$-PSR of a connected graph $G$ can be constructed with high probability in $R$ rounds with local capacity $O(1)$ and global capacity $\Gamma$.
    Then an \textsf{ST} and a \textsf{WFT} of $G$ can be constructed with high probability in $R+O(\log n)$ rounds with local capacity $O(1)$ and global capacity $O(\Gamma+\Delta\log n)$.
\end{lemma}
\begin{proof}
    Let $(H,I,A)$ be the constructed $\Delta$-PSR.
    Since $H$ is realized, connected, and of maximum degree at most $\Delta$, Theorem~\ref{thm:known-tree-construction} with $k=\Delta$ constructs a rooted spanning tree of $H$ and a \textsf{WFT} on $V$ in $O(\log n)$ rounds in $(0,\Delta\log n)$-\HYBRIDz w.h.p.
    The \textsf{WFT} is already a \textsf{WFT} of $G$ because $V(H)=V$.
    Applying $A$ to the spanning tree and $I$ yields an \textsf{ST} of $G$ in $O(\log n)$ further rounds in $(1,\Delta\log n)$-\HYBRIDz w.h.p.
    The three stages run one after another, so the round bounds add up and the capacity is the maximum of the three, and a union bound over the stages gives success w.h.p.
\end{proof}
Thus, it remains to construct an $O(1)$-PSR with standard capacity in $O(\log n)$ rounds w.h.p.
By Lemma~\ref{lem:pseudo-spanning-reduction}, such a construction proves Theorem~\ref{thm:low-capacity-st}.

\subsection{Input-edge forest, clusters, and edge responsibilities}
\label{subsec:clustering}
Toward constructing an $O(1)$-PSR, we first classify the connected components of the input graph by size.
The construction groups vertices into clusters of $\Theta(\log n)$ vertices, so components with fewer vertices are handled separately.
If $n=1$, we output the tree without edges.
Otherwise let $B\coloneqq\lceil\log n\rceil$, and call a connected component \emph{small} if it has fewer than $B$ vertices and \emph{large} otherwise.
In each large component, we construct a rooted input-edge forest $F$ whose trees, called \emph{$F$-components}, have size at least $B$ and height $O(\log n)$.
\begin{definition}[Cluster]
    A \emph{cluster} is a vertex subset $P$ of one $F$-component with $B\le|P|<4B$ and a \emph{controller} $c_P\in P$.
    The vertices of $P$ are its \emph{member vertices}.
    The cluster family $\mathcal P$ partitions the vertex set of each $F$-component.
    $P(v)$ denotes the cluster containing $v$, and $\ID(P)\coloneqq\ID(c_P)$.
\end{definition}
We construct $F$ and $\mathcal P$ in \S\ref{subsubsec:forest-and-clusters}.
We then select the representative input edges between clusters and assign responsibility for them to individual vertices in \S\ref{subsubsec:representative-selection}.

\subsubsection{Constructing the input-edge forest and clusters}
\label{subsubsec:forest-and-clusters}
\paragraph*{Input-edge forest}
We use the \textsf{MST} procedure of Garay, Kutten, and Peleg~\cite{garay1998sublinear}, which we call \textsf{GKP}, in Lenzen's formulation~\cite[Algorithm~16]{lenzen2019lecture}.
\textsf{GKP} merges the components of a growing forest of input edges as Bor\r{u}vka's algorithm does, except that in phase $i$ only the components of diameter at most $2^i$ select a minimum outgoing edge, which bounds the growth of the diameter.
After phase $i$ every component therefore has diameter $O(2^i)$ and, unless it spans its input connected component, size $\Omega(2^i)$~\cite[Lemmas~6.15 and~6.17]{lenzen2019lecture}.
We modify \textsf{GKP} by stopping a component once it reaches $B$ vertices.
A component $K$ has \emph{capped size} $\bar s(K)\coloneqq\min\{|K|,B\}$, and one with $\bar s(K)=B$ is \emph{saturated} and selects no further edge, while an \emph{active} component in phase $i$ is unsaturated, has diameter at most $2^i$, and has an edge leaving it.
A saturated component only stops selecting and still serves as an endpoint for edges selected by others, so after phase $i$ each tree has size at least $\min\{2^i,B\}$ unless it spans its input connected component.
Running $q\coloneqq\lceil\log B\rceil=\Theta(\log\log n)$ phases thus gives, in each large connected component, trees of size at least $B$ and height $O(\log n)$.
\textsf{GKP} runs in \CONGEST, which is equivalent to $(1,0)$-\HYBRIDz, and takes $O(2^i\log^*n)$ rounds in phase $i$~\cite[Corollary~6.16]{lenzen2019lecture}.
We replace every aggregation or broadcast inside an unsaturated component by $O(1)$ rounds of global communication and call the resulting procedure \emph{size-capped growth}, stated in full as Algorithm~\ref{alg:size-capped-growth} in Appendix~\ref{app:size-capped-growth-proof}.
An unsaturated component has fewer than $B$ vertices, so the following lemma provides this replacement.

\begin{lemma}
    \label{lem:representative-communication}
    Let $\mathcal S$ be a family of disjoint vertex sets of size $O(B)$, each with a representative $r_S\in S$ knowing and known by every $v\in S$.
    For all $S\in\mathcal S$ in parallel, $r_S$ and each $v\in S$ can exchange $O(1)$ messages in each direction in $O(1)$ rounds in $(0,\log n)$-\HYBRIDz.
\end{lemma}
\begin{proof}
    The global send count and the receive count are $O(B)=O(\log n)$ messages at each representative vertex and $O(1)$ messages at every other vertex.
    The sets are disjoint, and all destination IDs are known before the communication starts.
    The claim follows.
\end{proof}

\begin{lemma}
    \label{lem:size-capped-growth}
    Size-capped growth runs deterministically in $O(\log\log n\log^*n)$ rounds in $(1,\log n)$-\HYBRIDz.
    After phase $i$, each tree is rooted and has diameter $O(2^i)$, every vertex knows its root's ID and capped size, and the root of each unsaturated component stores the ID and parent ID of every member.
    Finally, each small component has one spanning tree, and each large component is partitioned into rooted trees of size $\ge B$ and height $O(\log n)$.
\end{lemma}
\begin{proof}[Proof sketch]
    The analysis of Lenzen~\cite[Lemmas~6.15 and~6.17]{lenzen2019lecture} applies with the size cap: after phase $i$, every tree has diameter $O(2^i)$ and, unless it spans its input connected component, size at least $\min\{2^i,B\}$.
    A phase of \textsf{GKP} is known to consist of $O(\log^*n)$ aggregations and broadcasts inside components~\cite[Corollary~6.16]{lenzen2019lecture}, and Lemma~\ref{lem:representative-communication} performs each of them in $O(1)$ rounds, with members filtering their reports to one item.
    The full proof is given in Appendix~\ref{app:size-capped-growth-proof}.
\end{proof}

\begin{figure}[htbp]
    \centering
    \includegraphics[width=0.93\textwidth]{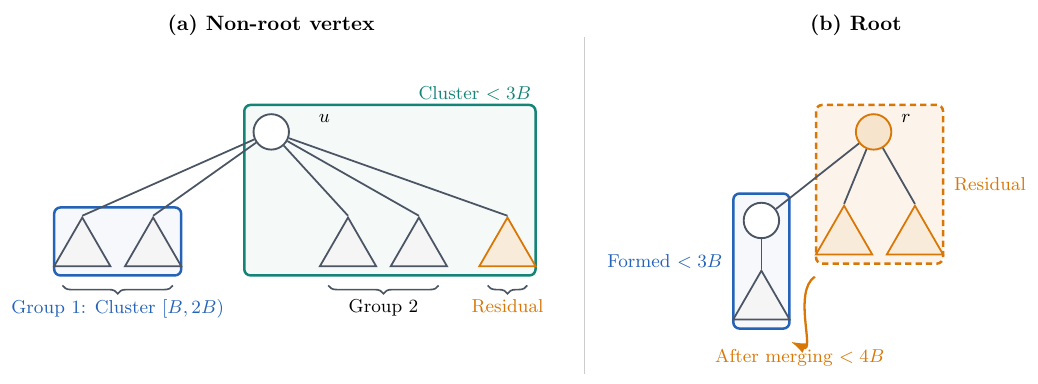}
    \caption{Bottom-up partition into clusters at an internal vertex (left) and at the root (right).}
    \label{fig:cluster-split}
\end{figure}

\paragraph*{Partitioning into clusters}
Figure~\ref{fig:cluster-split} illustrates the following bottom-up computation on the rooted tree $T$ of each $F$-component.
Let $U(u)$, the \emph{residual set} of $u$, be the set of vertices in the subtree of $u$ not yet assigned to a cluster after processing $u$, and let $s(u)\coloneqq|U(u)|$.
Vertex $u$ orders the nonempty residual sets of its children by ID and greedily closes a group of whole sets into a cluster as soon as the group reaches $B$ vertices, taking the child of its first set as controller.
Every residual set has size at most $B$, so each such cluster has size in $[B,2B)$.
Vertex $u$ then adds itself and the sets left over to one of the clusters it just closed.
If it closed none, these become the residual set $U(u)$ passed to its parent.
Each vertex sends $s(u)$ and one controller ID from its subtree, if any, to its parent.
Only the root $r$ has no parent, so a nonempty $U(r)$ joins a cluster named by a received controller ID, and the whole tree becomes one cluster with controller $r$ if no cluster exists.

\begin{lemma}
    \label{lem:cluster-construction}
    The above procedure partitions the vertex set of each tree of $F$ into clusters of size in $[B,4B)$ with pairwise distinct controllers.
    Each vertex knows its controller's ID, and each controller knows the IDs of all members of its cluster.
    It runs deterministically in $O(\log n)$ rounds in $(1,\log n)$-\HYBRIDz.
\end{lemma}
The size bounds follow from the grouping rule and the round bound from the $O(\log n)$ height of the trees of $F$ (Lemma~\ref{lem:size-capped-growth}).
Appendix~\ref{app:cluster-construction-proof} gives the proof.

\subsubsection{Selecting and assigning representative input edges}
\label{subsubsec:representative-selection}
We contract each cluster of \S\ref{subsubsec:forest-and-clusters} to a single vertex and run MPX-type clustering on the resulting graph.
For each selected input edge $e$, exactly one of the two clusters it joins becomes responsible for $e$ and hands it to one of its members, the \emph{responsible vertex} of $e$, which represents $e$ in the virtual network of \S\ref{subsec:arbitrary-degree-tree} in place of its endpoints.
By~\cite[Lemma~4.6]{gotte2023time}, each cluster becomes responsible for $O(\log n)$ input edges w.h.p., so the balancing leaves $O(1)$ edges at each of its $\Theta(\log n)$ members.
We use these edges to connect distinct $F$-components and discard those within one $F$-component.

Consider the simple graph on $\mathcal P$ in which two clusters are adjacent when at least one input edge joins them.
The MPX-type clustering of G\"{o}tte et al.~\cite[Lemmas~4.2 and~4.6]{gotte2023time} on this graph partitions $\mathcal P$ into parts, each with a rooted spanning tree.
We call these parts \emph{MPX components}, and the ID of the root cluster serves as the ID of the component.
Every cluster selects one input edge to each adjacent MPX component other than its own, and every non-root cluster also selects one input edge to its parent cluster.
The selected edges joining distinct $F$-components are the \emph{representative input edges}.
The cluster that selected such an edge, or the one with the smaller ID if both did, is \emph{responsible} for it, and its endpoint in that cluster is the \emph{responsible-side endpoint}.
Let $G'$ be the multigraph on $\mathcal P$ whose edges are the representative input edges, with parallel edges kept.

\begin{lemma}
    \label{lem:mpx-sparsification}
    The following properties hold with high probability.
    \begin{enumerate}
        \item $F$ together with all representative input edges connects each input connected component.
        \item Each cluster $P$ is adjacent to $k_P=O(\log n)$ other MPX components and is responsible for at most $1+k_P$ edges of $G'$.
    \end{enumerate}
\end{lemma}
\begin{proof}
    The parent edges connect each MPX component and the edges between components preserve adjacency between components, so the selected edges connect each input component.
    Each selected edge within an $F$-component can be replaced by a path in $F$, and every other selected edge is a representative input edge, which proves (1).
    By G\"{o}tte et al.~\cite[Lemma~4.6]{gotte2023time}, $k_P=O(\log n)$ for all clusters w.h.p., and $P$ selects at most one parent edge and one edge per adjacent component.
\end{proof}

\begin{lemma}
    \label{lem:mpx-implementation}
    The above procedure can be implemented in $O(\log n)$ rounds in $(1,\log n)$-\HYBRIDz.
    With high probability, both endpoints of the representative input edge corresponding to each edge of $G'$ know that it was selected and which endpoint's cluster is responsible.
\end{lemma}
\begin{proof}[Proof sketch]
    Each controller acts for its cluster in the MPX procedure of G\"{o}tte et al.~\cite{gotte2023time}.
    Forwarding every message arriving at a cluster's input edges to its controller could exceed the receive capacity.
    Each member therefore forwards only the highest-priority message under the MPX comparison rule, which yields the same outcome.
    Each iteration takes $O(1)$ rounds by Lemma~\ref{lem:representative-communication}, so MPX takes $O(\log n)$ rounds~\cite[Lemma~4.2]{gotte2023time}.
    Edges to adjacent components are reported one per round and finish in $O(\log n)$ rounds by Lemma~\ref{lem:mpx-sparsification}(2).
    The endpoints of each input edge then exchange whether their clusters selected it.
    Every real vertex handles $O(\log n)$ global messages per round and $O(1)$ per input-edge direction.
    The full proof is given in Appendix~\ref{app:mpx-implementation-proof}.
\end{proof}

Finally, each cluster orders the representative input edges $e=\{u,y\}$ for which it is responsible by $(\ID(u),\ID(y))$, with the responsible-side endpoint $u$ first, and assigns them to its members so that the numbers of assigned edges differ by at most one.
Let $p_e$ be the responsible vertex of $e$.

\begin{lemma}
    \label{lem:edge-distribution}
    The assignment can be implemented in $O(1)$ rounds in $(1,\log n)$-\HYBRIDz, and with high probability each real vertex is responsible for $O(1)$ representative input edges.
    Each $p_e$ belongs to the cluster responsible for $e$, knows both endpoint IDs and the responsible side, and both endpoints know $\ID(p_e)$.
\end{lemma}
Each cluster spreads $O(\log n)$ edges over $\Theta(\log n)$ members, and every notification passes between a controller and its own members or along $e$ itself.
The proof is given in Appendix~\ref{app:edge-distribution-proof}.

The assignment also supports the repeated communication used in \S\ref{subsec:arbitrary-degree-tree}.
\begin{lemma}
    \label{lem:assigned-edge-communication}
    After a successful assignment, a constant number of messages in each direction between $p_e$ and each endpoint of $e=\{u,y\}$, for all representative input edges in parallel, can be exchanged in $O(1)$ rounds in $(1,\log n)$-\HYBRIDz, routed along $p_e\to u\to y$.
\end{lemma}
The proof is given in Appendix~\ref{app:edge-distribution-proof}.

\subsection{Construction of an \texorpdfstring{$O(1)$}{O(1)}-pseudo-spanning representation}
\label{subsec:arbitrary-degree-tree}

We describe the construction using auxiliary vertices maintained by real vertices.
A real vertex \emph{simulates} an auxiliary vertex by performing its computation and communication.
An auxiliary vertex representing part of the role of a real vertex $x$ is a \emph{copy} of $x$, and its simulator may differ from $x$.
An edge whose endpoints have the same simulator needs no communication, and every other edge is realized by global communication between simulators that know each other's IDs.
The messages of every simulated vertex count toward its simulator's send and receive counts.
We also use \textsf{WFT} for rooted trees on vertex sets containing simulated vertices, with the same constant-degree and $O(\log n)$-diameter requirements.
Each endpoint's simulator holds the vertex IDs and simulator IDs of both endpoints of every tree edge.

\subsubsection{The expanded graph with copies and paths}
\begin{figure}[htbp]
    \centering
    \includegraphics[width=0.99\textwidth]{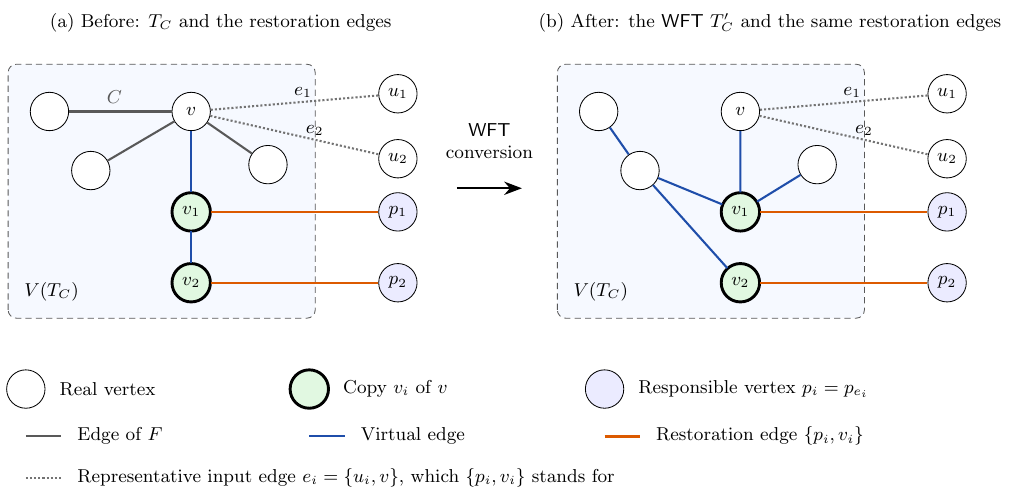}
    \caption{Construction of $\widehat H$ for one $F$-component $C$.
        The responsible vertex $p_i$ of $e_i=\{u_i,v\}$ holds a copy $v_i$ of $v$, and the restoration edge $\{p_i,v_i\}$ records the correspondence with $e_i$.
        The dotted $e_i$ are drawn for reference and are not edges of $\widehat H$.
        Transforming $T_C$ (left) into a \textsf{WFT} $T'_C$ (right) leaves the restoration edges unchanged.}
    \label{fig:pseudo-spanning-graph}
\end{figure}

Figure~\ref{fig:pseudo-spanning-graph} shows the construction for one large component $D$.
We build an \emph{expanded graph} $\widehat H$ in which copies are separate vertices, and obtain $H$ by identifying each vertex of $\widehat H$ with its simulator.
The edges of $F$ are not added to $\widehat H$ and are retained in the side information for restoration.
\begin{enumerate}
    \item For each $v$, let $e_1,\ldots,e_t$ with $e_i=\{u_i,v\}$ be the representative input edges incident to $v$ for which $u_i$ is the responsible-side endpoint, ordered by $\ID(u_i)$.
          Its responsible vertex $p_i\coloneqq p_{e_i}$ creates and simulates a copy $v_i$ of $v$, adds the \emph{restoration edge} $\{p_i,v_i\}$ to $\widehat H$, and records its correspondence with $e_i$.
    \item Let $T_C$ consist of an $F$-component $C$ and the paths $v,v_1,\ldots,v_t$ for $v\in V(C)$, rooted at the root of $C$.
          Keeping copies distinct, Lemma~\ref{lem:tree-to-wft} turns $T_C$ into a \textsf{WFT} $T'_C$ on $V(T_C)$, whose edges we add to $\widehat H$.
    \item Let $\sigma$ map each vertex of $\widehat H$ to its simulator, i.e., $\sigma(v)\coloneqq v$ and $\sigma(v_i)\coloneqq p_i$, and let $H$ be the graph on $V(D)$ with an edge $\{\sigma(a),\sigma(b)\}$ for every edge $\{a,b\}$ of $\widehat H$ with $\sigma(a)\ne\sigma(b)$, parallel edges merged.
          Edges whose endpoints have the same simulator, in particular all restoration edges, need no communication.
          The side information $I$ of a real vertex records the edges of $\widehat H$ at it and its copies, their opposite simulator IDs, its restoration-edge correspondences, and its incident edges of $F$ with their parent--child relations.
\end{enumerate}

\begin{lemma}
    \label{lem:constant-degree-cluster-overlay}
    The following properties hold with high probability.
    \begin{enumerate}
        \item Each connected component of $\widehat H$ minus the restoration edges is the \textsf{WFT} $T'_C$ of exactly one $F$-component $C$.
        \item Each restoration edge joins two distinct such components, and its representative input edge joins the corresponding $F$-components.
        \item Each real vertex simulates $O(1)$ copies, $\widehat H$ has maximum degree $O(1)$, and $H$ is a connected graph on $V(D)$ with maximum degree $O(1)$.
    \end{enumerate}
\end{lemma}
The proof is given in Appendix~\ref{app:constant-degree-cluster-overlay-proof}.

\begin{lemma}
    \label{lem:overlay-realization}
    The construction runs in $O(\log n)$ rounds in $(1,\log n)$-\HYBRIDz and, w.h.p., realizes $H$ and constructs $I$.
    For each edge of $\widehat H$, the endpoint simulators know each other's simulator ID and both endpoint vertex IDs, and $p_i$ retains the endpoint IDs of $e_i$.
\end{lemma}
\begin{proof}[Proof sketch]
    A copy ID is an $O(\log n)$-bit tuple of a tag, $\ID(p_i)$, and the endpoint IDs of $e_i$, which $v$ can compute.
    $v$ sends each copy the IDs of its path neighbors via $v\to u_i\to p_i$ in $O(1)$ rounds (Lemma~\ref{lem:assigned-edge-communication}).
    The only virtual edges of $T_C$ are at most two path edges per vertex, so Lemma~\ref{lem:tree-to-wft} with $d=2$ transforms all $T_C$ in parallel.
    Each real vertex simulates $O(1)$ vertices, so its message counts grow by a constant factor.
    The full proof is given in Appendix~\ref{app:overlay-realization-proof}.
\end{proof}

\subsubsection{Restoration of an \textsf{ST}}
Algorithm~\ref{alg:psr-restoration} restores an \textsf{ST} by expanding $T$ into a spanning tree $\widehat T$ of $\widehat H$, applying quotient-tree selection to the components $V(T'_C)$, and replacing the selected restoration edges by their representative input edges (edges are compared lexicographically by endpoint IDs).
\begin{algorithm}[H]
    \caption{Restoration algorithm $A$ (given a rooted spanning tree $T$ of $H$ and $I$)}
    \label{alg:psr-restoration}
    \begin{algorithmic}[1]
        \State Replace each edge of $T$ by the smallest corresponding edge of $\widehat H$, and let $\widehat T$ be these edges with all restoration edges
        \State Root $\widehat T$ at the root $r$ of $T$ and compute all depths
        \State In each component $V(T'_C)$ of $\widehat H$ minus the restoration edges, find the vertex minimizing the pair of depth and vertex ID
        \State For each component without $r$, replace the parent edge of its chosen vertex by the corresponding representative input edge
        \State Reroot the selected representative input edges together with $F$ at $r$ and output the result
    \end{algorithmic}
\end{algorithm}

\begin{lemma}
    \label{lem:st-restoration}
    For $H$ and $I$ obtained by a successful construction, Algorithm~\ref{alg:psr-restoration} restores an \textsf{ST} of $D$ from any rooted spanning tree of $H$, preserving the root, deterministically in $O(\log n)$ rounds in $(1,\log n)$-\HYBRIDz.
\end{lemma}
\begin{proof}
    The restoration edges join each real vertex $x$ to every copy it simulates, so the vertices with simulator $x$ form a star centered at $x$.
    The replacement edges connect these stars according to $T$, so $\widehat T$ is a spanning tree of $\widehat H$.
    Lines 3--4 apply Lemma~\ref{lem:quotient-tree-selection} to $\widehat T$ with the partition of Lemma~\ref{lem:constant-degree-cluster-overlay}(1).
    Only restoration edges cross between the parts, so all selected edges are restoration edges, and they form a spanning tree of the contracted graph, i.e., of the $F$-components.
    By Lemma~\ref{lem:constant-degree-cluster-overlay}(2), replacing each selected $\{p_i,v_i\}$ by $\{u_i,v\}$ preserves these connections, and adding $F$ gives a connected graph with $(|V(D)|-m)+(m-1)=|V(D)|-1$ edges for $m$ $F$-components, hence an input-edge spanning tree of $D$.
    The round and capacity bounds are verified in Appendix~\ref{app:st-restoration-bounds}.
\end{proof}

\begin{proof}[Proof of Theorem~\ref{thm:low-capacity-st}]
    Apply Lemma~\ref{lem:size-capped-growth} to all components.
    A small component already has a spanning tree, whose root informs all vertices by Lemma~\ref{lem:representative-communication} and which Lemma~\ref{lem:tree-to-wft} turns into a \textsf{WFT}.
    In a large component, Lemmas~\ref{lem:constant-degree-cluster-overlay}(3), \ref{lem:overlay-realization}, and~\ref{lem:st-restoration} make $(H,I,A)$ an $O(1)$-PSR built in $O(\log n)$ rounds with standard capacity, so Lemma~\ref{lem:pseudo-spanning-reduction} applies.
    Sufficiently large constants and a union bound over the at most $n$ components give success w.h.p.
\end{proof}

\section{Few-weight \textsf{MST} and \texorpdfstring{$(1+\epsilon)$}{(1+epsilon)}-approximate \textsf{MST}}
\label{sec:kweight-mst}

\subsection{A general reduction from \textsf{ST} construction}
\label{subsec:kweight-reduction}
We formulate the threshold-subgraph construction of Gmyr et al.~\cite{gmyr2017distributed} as a reduction to \textsf{ST} construction.
Let the distinct weights be $w_1<\cdots<w_k$, and let $G_i=(V,\{e:w(e)\le w_i\})$ and $G_0=(V,\emptyset)$.
To process $w_i$, we apply the assumed \textsf{ST} algorithm to $G_i$ and to $G_{i-1}$.
Let $T_{i,D}$ be the tree of a $G_i$-component $D$, whose depths we compute by Lemma~\ref{lem:fhs-tree-rooting}.
We apply quotient-tree selection to $T_{i,D}$ with the partition of $D$ into $G_{i-1}$-components.
Within each part, we aggregate on the \textsf{WFT} obtained from its tree by Lemma~\ref{lem:tree-to-wft}.
Let $A_i$ be the set of selected edges.
We notify both endpoints of each selected edge over that edge.
For parallelism $p$, an \textsf{ST} of $G$ and its \textsf{WFT} disseminate the weights in batches of $p$.
Each vertex merges its incident weights with the lists of its children, removes weights already shared, and sends the $p$ smallest to its parent, which the root then broadcasts.
We stop after a batch of fewer than $p$ weights.
Each batch is processed in parallel, and the output is $\bigcup_i A_i$.

\begin{proof}[Proof of Theorem~\ref{thm:kweight-mst}]
    The $G_{i-1}$-components partition $V(T_{i,D})$, so by Lemma~\ref{lem:quotient-tree-selection}, $A_i$ connects them within $D$ as a tree, and each selected edge lies in $G_i$ and joins distinct $G_{i-1}$-components, hence has weight $w_i$.
    Let $B_i\coloneqq\bigcup_{j\le i}A_j$.
    Inductively, $(V,B_i)$ is a spanning forest of $G_i$.
    Thus, $T\coloneqq(V,B_k)$ is a spanning tree with exactly $n-\kappa_i$ edges of weight at most $w_i$, where $\kappa_i$ is the number of components of $G_i$.
    Any spanning tree has at most $n-\kappa_i$ such edges, since they form a forest in $G_i$.
    Thus, the $j$th smallest weight in $T$ is at most that in any other spanning tree.
    Therefore, $T$ is an \textsf{MST}.
    The round and capacity bounds are verified in Appendix~\ref{app:kweight-analysis}.
\end{proof}

Substituting Theorem~\ref{thm:low-capacity-st} gives Corollary~\ref{cor:kweight-mst}.
If the input graph is a realized graph of constant degree, all local communication in the reduction becomes global, giving the same bound in $(0,p\log n)$-\HYBRIDz (Corollary~\ref{cor:constant-degree-kweight-mst} in Appendix~\ref{app:constant-degree-mst}).

\subsection{Geometric rounding for \texorpdfstring{$(1+\epsilon)$}{(1+epsilon)}-approximate \textsf{MST}}
Following Gmyr et al.~\cite[Theorem~17]{gmyr2017distributed}, we round positive weights up to powers of $1+\epsilon$, setting $\ell(e)\coloneqq0$ if $w(e)=0$ and $\ell(e)\coloneqq(1+\epsilon)^{\lceil\log_{1+\epsilon}w(e)\rceil}$ otherwise.
Edges with the same $\ell(e)$ form a \emph{class}, and $\widehat h_\epsilon\coloneqq\max\{1,|\{\ell(e):e\in E\}|\}$ is the number of classes.
\begin{proof}[Proof sketch of Corollary~\ref{cor:approx-mst}]
    Since $w(e)\le\ell(e)\le(1+\epsilon)w(e)$, an \textsf{MST} for $\ell$ is a $(1+\epsilon)$-approximate \textsf{MST} for $w$, and each class is represented during weight sharing by its smallest input weight, an $O(\log n)$-bit value preserving the order of classes.
    Corollary~\ref{cor:kweight-mst} for $\ell$ with parallelism $p'\coloneqq\min\{p,\widehat h_\epsilon\}$ takes $O((\widehat h_\epsilon/p')\log n)$ rounds in $(p',p'\log n)$-\HYBRIDz, and $\widehat h_\epsilon\le2+\lceil\log_{1+\epsilon}W\rceil=O(\epsilon^{-1}\log W)$ for $0<\epsilon\le1$.
    The full proof is given in Appendix~\ref{app:approx-mst-proof}.
\end{proof}

\section{Application to biconnected components}
\label{sec:bcc}
A connected graph with at least three vertices is \emph{biconnected} if removing any vertex leaves it connected, a maximal biconnected subgraph is a \emph{biconnected component} (BCC), and a bridge is a degenerate BCC with one edge.
Every input edge belongs to exactly one BCC, and a vertex is an articulation point if and only if its incident edges belong to at least two BCCs.

We follow the implementation of the Tarjan--Vishkin reduction~\cite{tarjan1985efficient} by G\"{o}tte et al.~\cite{gotte2023time}, which recovers the BCCs from the components of an auxiliary graph $H_{\mathrm{BCC}}$ on the tree edges of a rooted \textsf{ST}, simulating each auxiliary edge over an input edge.
Hence $H_{\mathrm{BCC}}$, after adding isolated roots, is isomorphic to a spanning subgraph of the input graph.
Theorem~\ref{thm:low-capacity-st} replaces their \textsf{ST} and \textsf{WFT} constructions, and an Euler-tour aggregation in which each vertex holds $O(1)$ items replaces their aggregation of subtree values through multicast trees, the remaining step that exceeds standard capacity.
In the Euler tour of the \textsf{ST}, each non-root vertex $v$ with parent $p(v)$ holds the two items $(p(v),v)$ and $(v,p(v))$, and consecutive items are held by the same vertex, by tree neighbors, or by siblings, whose IDs the parent introduces over input edges.
Pointer jumping on this list computes prefix sums, which give the preorder numbers and subtree sizes, and the minimum and maximum over the interval between the two items of $v$, which covers exactly the subtree of $v$.
Each vertex sends and receives $O(1)$ global messages per round.
Appendix~\ref{app:bcc} gives the auxiliary graph, the Euler-tour preprocessing, and the proof of Theorem~\ref{thm:biconnectivity}.

\section{Conclusion}
\label{sec:conclusion}
We constructed an $O(1)$-pseudo-spanning representation, which yields a \textsf{WFT} and an \textsf{ST} in $O(\log n)$ rounds w.h.p.\ with standard capacity, and applied it to few-weight and approximate \textsf{MST} and to biconnected components.
Since constructing a \textsf{WFT} requires $\Omega(\log n)$ rounds~\cite{gotte2023time}, our \textsf{WFT} construction is round-optimal.
For \textsf{ST} and \textsf{MST}, we are not aware of stronger lower bounds.
The sparse-cut bottlenecks behind the two-party reductions that are standard in \CONGEST~\cite{peleg2000near} do not carry over directly, because global communication can carry $\Omega(n\log n)$ messages per round across a balanced cut that few input edges cross.
Whether exact \textsf{MST} with arbitrary weights can be computed in $O(\log n)$ rounds with standard capacity remains open.
Our construction is randomized through the MPX-type clustering and the constant-degree tree construction (Theorem~\ref{thm:known-tree-construction}), and the best known deterministic bound is $O(\log^2 n)$~\cite{feldmann2021fast,gmyr2017distributed}.
Achieving the same bounds deterministically, or together with $O(n\operatorname{polylog}n)$ total messages~\cite{chang2025overlay}, also remains open.

\paragraph*{Acknowledgments.}
Shoma Hiraoka was supported by JST BOOST, Japan Grant Number JPMJBS2402.
Naoki Kitamura was supported by JSPS KAKENHI Grant Numbers JP23K16838 and JP26K21168.
Taisuke Izumi was supported by JSPS KAKENHI Grant Number JP23K24825.
This work was also supported by JST CRONOS, Japan Grant Number JPMJCS24K2 and JST ASPIRE, Japan Grant Number JPMJAP2302.

\paragraph*{Use of AI tools.}
We used GPT-5.6 Sol and GPT-6 Astra (OpenAI) and Claude Opus 5.5 and Fable 5.2 (Anthropic) for writing assistance, verified all AI-assisted content, and take full responsibility.

\bibliographystyle{plainurl}
\bibliography{references}

\begin{thebibliography}{10}

\bibitem{angluin2005fast}
Dana Angluin, James Aspnes, Jiang Chen, Yinghua Wu, and Yitong Yin.
\newblock Fast construction of overlay networks.
\newblock In {\em Proceedings of the Seventeenth Annual ACM Symposium on
  Parallelism in Algorithms and Architectures}, pages 145--154. Association for
  Computing Machinery, 2005.
\newblock \href {https://doi.org/10.1145/1073970.1073991}
  {\path{doi:10.1145/1073970.1073991}}.

\bibitem{augustine2019distributed}
John Augustine, Mohsen Ghaffari, Robert Gmyr, Kristian Hinnenthal, Fabian Kuhn,
  Jason Li, and Christian Scheideler.
\newblock Distributed computation in node-capacitated networks.
\newblock In {\em Proceedings of the 31st ACM Symposium on Parallelism in
  Algorithms and Architectures}, pages 69--79. Association for Computing
  Machinery, 2019.
\newblock \href {https://doi.org/10.1145/3323165.3323195}
  {\path{doi:10.1145/3323165.3323195}}.

\bibitem{augustine2020shortest}
John Augustine, Kristian Hinnenthal, Fabian Kuhn, Christian Scheideler, and
  Philipp Schneider.
\newblock Shortest paths in a {HYBRID} network model.
\newblock In {\em Proceedings of the 2020 ACM-SIAM Symposium on Discrete
  Algorithms}, pages 1280--1299. Society for Industrial and Applied
  Mathematics, 2020.
\newblock \href {https://doi.org/10.1137/1.9781611975994.78}
  {\path{doi:10.1137/1.9781611975994.78}}.

\bibitem{chang2025overlay}
Yi-Jun Chang, Yanyu Chen, and Gopinath Mishra.
\newblock Overlay network construction: Improved overall and node-wise message
  complexity.
\newblock In {\em 45th IARCS Annual Conference on Foundations of Software
  Technology and Theoretical Computer Science (FSTTCS 2025)}, volume 360 of
  {\em Leibniz International Proceedings in Informatics (LIPIcs)}, pages
  21:1--21:21. Schloss Dagstuhl--Leibniz-Zentrum f{\"u}r Informatik, 2025.
\newblock Full version: arXiv:2412.04771. Appendix references follow the full
  version.
\newblock \href {https://doi.org/10.4230/LIPIcs.FSTTCS.2025.21}
  {\path{doi:10.4230/LIPIcs.FSTTCS.2025.21}}.

\bibitem{feldmann2021fast}
Michael Feldmann, Kristian Hinnenthal, and Christian Scheideler.
\newblock Fast {HYBRID} network algorithms for shortest paths in sparse graphs.
\newblock In {\em 24th International Conference on Principles of Distributed
  Systems (OPODIS 2020)}, volume 184 of {\em Leibniz International Proceedings
  in Informatics (LIPIcs)}, pages 31:1--31:16. Schloss
  Dagstuhl--Leibniz-Zentrum f{\"u}r Informatik, 2021.
\newblock \href {https://doi.org/10.4230/LIPIcs.OPODIS.2020.31}
  {\path{doi:10.4230/LIPIcs.OPODIS.2020.31}}.

\bibitem{fraigniaud2022computing}
Pierre Fraigniaud, Pedro Montealegre, Pablo Paredes, Ivan Rapaport,
  Mart{\'{\i}}n R{\'{\i}}os-Wilson, and Ioan Todinca.
\newblock Computing power of hybrid models in synchronous networks.
\newblock In {\em 26th International Conference on Principles of Distributed
  Systems (OPODIS 2022)}, volume 253 of {\em Leibniz International Proceedings
  in Informatics (LIPIcs)}, pages 20:1--20:18. Schloss
  Dagstuhl--Leibniz-Zentrum f{\"u}r Informatik, 2022.
\newblock \href {https://doi.org/10.4230/LIPIcs.OPODIS.2022.20}
  {\path{doi:10.4230/LIPIcs.OPODIS.2022.20}}.

\bibitem{garay1998sublinear}
Juan~A. Garay, Shay Kutten, and David Peleg.
\newblock A sublinear time distributed algorithm for minimum-weight spanning
  trees.
\newblock {\em SIAM Journal on Computing}, 27(1):302--316, 1998.
\newblock \href {https://doi.org/10.1137/S0097539794261118}
  {\path{doi:10.1137/S0097539794261118}}.

\bibitem{gmyr2017distributed}
Robert Gmyr, Kristian Hinnenthal, Christian Scheideler, and Christian Sohler.
\newblock Distributed monitoring of network properties: The power of {HYBRID}
  networks.
\newblock In {\em 44th International Colloquium on Automata, Languages, and
  Programming (ICALP 2017)}, volume~80 of {\em Leibniz International
  Proceedings in Informatics (LIPIcs)}, pages 137:1--137:15. Schloss
  Dagstuhl--Leibniz-Zentrum f{\"u}r Informatik, 2017.
\newblock \href {https://doi.org/10.4230/LIPIcs.ICALP.2017.137}
  {\path{doi:10.4230/LIPIcs.ICALP.2017.137}}.

\bibitem{gotte2023time}
Thorsten G{\"o}tte, Kristian Hinnenthal, Christian Scheideler, and Julian
  Werthmann.
\newblock Time-optimal construction of overlay networks.
\newblock {\em Distributed Computing}, 36(3):313--347, 2023.
\newblock \href {https://doi.org/10.1007/s00446-023-00442-4}
  {\path{doi:10.1007/s00446-023-00442-4}}.

\bibitem{lenzen2019lecture}
Christoph Lenzen.
\newblock Lecture notes on theory of distributed systems, 2019.
\newblock Lecture notes.
\newblock URL:
  \url{https://www.mpi-inf.mpg.de/fileadmin/inf/d1/teaching/winter19/tods/ToDS.pdf}.

\bibitem{miller2013parallel}
Gary~L. Miller, Richard Peng, and Shen~Chen Xu.
\newblock Parallel graph decompositions using random shifts.
\newblock In {\em Proceedings of the 25th ACM Symposium on Parallelism in
  Algorithms and Architectures}, pages 196--203. Association for Computing
  Machinery, 2013.
\newblock \href {https://doi.org/10.1145/2486159.2486180}
  {\path{doi:10.1145/2486159.2486180}}.

\bibitem{nowicki2018random}
Krzysztof Nowicki.
\newblock Random sampling applied to the {MSF} problem in the node congested
  clique model, 2018.
\newblock \href {https://arxiv.org/abs/1807.08738} {\path{arXiv:1807.08738}}.

\bibitem{peleg2000near}
David Peleg and Vitaly Rubinovich.
\newblock A near-tight lower bound on the time complexity of distributed
  minimum-weight spanning tree construction.
\newblock {\em SIAM Journal on Computing}, 30(5):1427--1442, 2000.
\newblock \href {https://doi.org/10.1137/S0097539700369740}
  {\path{doi:10.1137/S0097539700369740}}.

\bibitem{tarjan1985efficient}
Robert~E. Tarjan and Uzi Vishkin.
\newblock An efficient parallel biconnectivity algorithm.
\newblock {\em SIAM Journal on Computing}, 14(4):862--874, 1985.
\newblock \href {https://doi.org/10.1137/0214061} {\path{doi:10.1137/0214061}}.

\end{thebibliography}

\clearpage
\appendix
\section{Details of Theorem~\ref{thm:known-tree-construction}}
\label{app:known-tree-construction-details}
We use the copies and their simulation described in Section~\ref{subsec:arbitrary-degree-tree}.
As preprocessing, G\"{o}tte et al.\ replace each vertex by one copy per incident edge and connect its copies in a cycle to reduce the maximum degree to 3~\cite{gotte2023time}.
We instead connect the copies in a path and restore a spanning tree of the original graph by quotient-tree selection (Algorithm~\ref{alg:degree-reduction}).
We write $d_H(x)$ for the degree of $x$ in $H$.
\begin{algorithm}[H]
    \caption{Degree reduction and spanning-tree restoration for a realized graph $H$}
    \label{alg:degree-reduction}
    \begin{algorithmic}[1]
        \State Each $x\in V(H)$ orders its incident edges as $e_{x,1},\ldots,e_{x,d_H(x)}$ and creates a copy $x_i$ for each $e_{x,i}$ (if $d_H(x)=0$, it creates one copy $x_1$)
        \State For each $x\in V(H)$, connect its copies in index order into a path $P_x$
        \State For each edge $e=\{x,y\}=e_{x,i}=e_{y,j}$, connect $x_i$ and $y_j$, giving the graph $\widetilde H$
        \State Compute a spanning tree $\widetilde T$ of $\widetilde H$ by the constant-degree algorithm of Feldmann et al.
        \State Apply Lemma~\ref{lem:fhs-tree-rooting} to root $\widetilde T$ (keeping its root if already rooted) and compute the depth of each copy
        \State Each $x\in V(H)$ selects the copy $z_x$ on $P_x$ with the minimum pair of depth and copy ID
        \State For each $x$ other than the vertex represented by the root, retain the edge of $H$ corresponding to the parent edge of $z_x$ as the parent edge of $x$
    \end{algorithmic}
\end{algorithm}

\begin{lemma}
    \label{lem:degree-reduction}
    Let $H$ be a realized connected graph on a set of real vertices, and let $\widetilde H$ be the graph obtained in lines 1--3 of Algorithm~\ref{alg:degree-reduction}.
    The following properties hold.
    \begin{enumerate}
        \item $\widetilde H$ is connected and has maximum degree at most 3.
              Each real vertex $x$ simulates its $\max\{1,d_H(x)\}$ copies, and exchanging one notification in each direction of every edge of $H$ gives the simulators of each edge of $\widetilde H$ the vertex IDs and simulator IDs of both endpoints.
        \item For any rooted spanning tree $\widetilde T$ of $\widetilde H$, the edges selected by applying lines 5--7 to it form a rooted spanning tree of $H$ whose root is the vertex represented by the root of $\widetilde T$.
    \end{enumerate}
\end{lemma}
\begin{proof}
    Each copy is incident to at most two path edges and one original edge, so the maximum degree is at most 3.
    Each path $P_x$ connects the copies of $x$, and each original edge connects its corresponding copies, so the connectivity of $H$ implies that of $\widetilde H$.
    The ID of a copy $x_i$ is determined by the ID of $x$ and its local index $i$, and $x$ simulates all copies on $P_x$.
    The endpoints of each path edge are simulated by the same real vertex and require no communication.
    If the endpoints of each original edge $\{x,y\}$ exchange their corresponding copy IDs, the endpoints of each edge of $\widetilde H$ know each other's IDs.

    To prove the second property, we apply Lemma~\ref{lem:quotient-tree-selection} to $\widetilde T$ with the partition $\{V(P_x)\}_{x\in V(H)}$ of $V(\widetilde H)$.
    Contracting each $P_x$ to one vertex gives exactly $H$, and each edge between paths corresponds to an original edge of $H$, so the selected edges form a rooted spanning tree of $H$.
\end{proof}

\begin{proof}[Proof of Theorem~\ref{thm:known-tree-construction}]
    The case $|U|=1$ is trivial.
    We construct the graph $\widetilde H$ of maximum degree 3 by lines 1--3 of Algorithm~\ref{alg:degree-reduction} and send the notifications of Lemma~\ref{lem:degree-reduction}(1) by global communication along the edges of $H$.
    On the tree obtained in line 4, the aggregation algorithm of Feldmann et al.~\cite[Lemma~3]{feldmann2021fast} computes and distributes the smallest simulator ID $\ID(r)$.
    Choosing the first copy $r_1$ of $r$ as the root in line 5 yields a spanning tree $T$ of $H$ rooted at $r$, by Lemma~\ref{lem:degree-reduction}(2).
    Each simulator $x$ performs the selection within its path $P_x$ by local computation.
    We report the selected edges and parent--child relations to both endpoints of the corresponding edges of $H$.

    Next, we treat all edges of $T$ as virtual edges and apply Lemma~\ref{lem:tree-to-wft} with $d=k$, using the all-virtual variant of Remark~\ref{rem:tree-to-wft-variants}.
    This yields a \textsf{WFT} on $U$ in $O(\log n)$ rounds with global capacity $O(k+\log n)$ and no local communication.

    During the operations on $\widetilde H$, each real vertex simulates $O(k)$ copies.
    Spanning-tree construction, aggregation, and rooting each use $O(\log n)$ global sends and receives per copy per round, giving $O(k\log n)$ of each per real vertex.
    Notifications of copy IDs and edge selection use known neighbor IDs in $H$ and require $O(k)$ sends and receives per real vertex.
    Messages include copy IDs and simulator IDs.
    Since $k\ge1$, the \textsf{WFT} transformation also fits the stated capacity, and the total number of rounds is $O(\log n)$.
    The graph $\widetilde H$ has $O(nk)=O(n^2)$ vertices, so its IDs fit in $O(\log n)$ bits.
    We set the parameters of randomized spanning-tree construction to obtain a high-probability guarantee with respect to the global input size $n$.
    All remaining operations are deterministic.
\end{proof}

\section{Input-edge forests and clusters}
\label{app:forest-clusters}

\subsection{Algorithm~\ref{alg:size-capped-growth} and proof of Lemma~\ref{lem:size-capped-growth}}
\label{app:size-capped-growth-proof}
We first fix the terminology used by the algorithm and its analysis.
Let $\mathcal F$ be the edge set of the forest built so far.
A \emph{tree component} is a connected component of $(V,\mathcal F)$, and an input edge is \emph{outgoing} for a tree component if it joins that component to another one.
Outgoing edges are compared by weight and then by endpoint IDs, so each tree component has a unique minimum outgoing input edge.
Capped size, saturation, and activity are as defined in \S\ref{subsubsec:forest-and-clusters}.
A saturated tree component keeps its root and only serves as an endpoint for edges selected by others, so its size may exceed $B$.
Adding every selected edge at once could chain many tree components together and increase the diameter.
The algorithm therefore adds a maximal matching of the selected edges first, and then the selected edge of each unmatched active component.
\begin{algorithm}[H]
    \caption{Size-capped growth of tree components.
        Bold text indicates the iteration range and merging conditions modified from~\cite[Algorithm~16]{lenzen2019lecture}.}
    \label{alg:size-capped-growth}
    \begin{algorithmic}[1]
        \State $\mathcal F\coloneqq\emptyset$
        \For{$\boldsymbol{i=1,\ldots,\lceil\log B\rceil}$}
        \State For each tree component $K$, mark $K$ as active if and only if its diameter is at most $2^i$, $\boldsymbol{\bar s(K)<B}$, and it has an outgoing input edge
        \State Each active component $K$ selects its minimum outgoing input edge
        \State Find a maximal matching $M$ in the graph whose vertices are tree components and whose edges are the selected edges \textbf{with both tree components active}, and add it to $\mathcal F$
        \State Each active component not incident to $M$ adds its selected edge to $\mathcal F$
        \EndFor
    \end{algorithmic}
\end{algorithm}

\begin{proof}[Proof of Lemma~\ref{lem:size-capped-growth}]
    The edges selected by active components form a forest in the graph obtained by contracting each tree component to one vertex.
    Otherwise, the component that selected the maximum edge of a cycle would not have selected its minimum outgoing edge.
    By maximality of the matching, each unmatched active component connects to a matched component or an inactive component.
    Consider a tree formed by merging at least two tree components, and contract each previous tree component to one vertex.
    The result is a tree obtained by attaching active components to the two endpoints of one matching edge, or a star centered at an inactive component.
    Thus, the contracted tree has diameter at most three and contains at most one inactive component.
    As in Lenzen's diameter analysis~\cite[Lemma~6.15]{lenzen2019lecture}, the maximum diameter $D_i$ after phase $i$ satisfies
    \[
        D_0=0,\qquad D_i\le D_{i-1}+3\cdot2^i+3
    \]
    and hence $D_i=O(2^i)$.
    Since $2^q=O(B)$, the final height is $O(\log n)$.

    Following Lenzen's induction~\cite[Proof of Lemma~6.17]{lenzen2019lecture}, we show that after phase $i$, each tree that does not yet span its input connected component has size at least $t_i\coloneqq\min\{2^i,B\}$.
    The case $i=0$ is immediate.
    Suppose that a tree violating this lower bound remains after phase $i$.
    All previous tree components contained in it are unsaturated, have diameter less than $2^i$, and have outgoing edges, so they are active.
    At least two therefore merge, and the induction hypothesis gives a size of at least $2t_{i-1}\ge t_i$, a contradiction.
    Since $t_q=B$, small connected components have spanning trees, and each tree in a large connected component has size at least $B$.

    Initially, every tree is a singleton, so its root information and each vertex's root ID and capped size are known.
    For the implementation, we use Lenzen's procedures for edge selection and maximal matching~\cite[Lemma~6.11 and Corollary~6.16]{lenzen2019lecture}.
    Within an unsaturated tree component, each aggregation or broadcast required by these procedures takes a constant number of rounds by Lemma~\ref{lem:representative-communication}.
    Communication between components uses input edges.
    Each member vertex filters its report of the minimum outgoing edge or its matching proposals to one item before sending it to the root, so each root receives $O(B)$ messages per round.
    We orient each selected edge between active components away from the side that selected it, breaking ties by root IDs when both sides selected the same edge.
    This gives a forest of out-degree at most one.
    We simulate Lenzen's color-reduction procedure on this forest to obtain a maximal matching in $O(\log^*n)$ rounds.
    In each matching iteration, both components learn whether the edge is accepted, and both input-edge endpoints also learn which edges are added to the forest.

    When at least two tree components merge, the new root is the root of the inactive component, if one exists.
    Otherwise, it is the root with the smaller ID among the two components joined by the matching edge.
    Let $K^\star$ be the previous component containing this root.
    We root the contracted tree at $K^\star$.
    Every other component is unsaturated, so its root is informed of its endpoint of the edge to the parent component and uses the stored parent information to reroot its component at that endpoint.
    Each vertex learns its new parent, and the connection endpoint takes the opposite endpoint as its parent.

    If $K^\star$ was unsaturated before the merge, we aggregate the component sizes on the contracted tree, capping the value at $B$ after each addition at a member vertex or a root.
    Each member vertex sends only one value to its root, so the root receives $O(B)$ messages even if many components connect to it.
    If $K^\star$ is saturated, we omit the aggregation because the capped size after the merge is $B$.
    We broadcast the new root ID and the capped size, omitting the broadcast within a saturated component because both values remain unchanged there.
    If the merged component is unsaturated, each vertex sends its own ID and its updated parent's ID to the new root, starting no earlier than the round after receiving the root ID.
    This preserves the information stated in the lemma for the next phase.

    The contracted tree has diameter at most three, so the updates after a merge take a constant number of rounds.
    At each real vertex, the global send count and the receive count are $O(B)$ messages per round, and local capacity $O(1)$ suffices.
    Hence, one phase takes $O(\log^*n)$ rounds, and all $q$ phases take $O(\log\log n\log^*n)$ rounds.
\end{proof}

\subsection{Proof of Lemma~\ref{lem:cluster-construction}}
\label{app:cluster-construction-proof}
\begin{proof}[Proof of Lemma~\ref{lem:cluster-construction}]
    No residual set of a child is split, so each vertex has a unique assignment.
    A vertex that closes no group leaves the residual sets of all of its children ungrouped, and their total is then less than $B$, so its own residual set has size at most $B$.
    A group closes as soon as its total reaches $B$ and is enlarged by whole residual sets of size at most $B$, so it has size in $[B,2B)$.
    At most $B$ further vertices are added when the cluster is finalized and again in the final step at the root, so the final size lies in $[B,4B)$.
    If no cluster exists, then $U(r)=V(T)$, and $|T|\ge B$ together with $|U(r)|\le B$ gives $|T|=B$.
    Each controller belongs to its own cluster, so the controllers are pairwise distinct.

    During the bottom-up pass, each vertex records the cluster to which it added each residual set received from a child, or records that it left the set for its parent to process.
    Once the root's residual set is assigned, each vertex sends the controller IDs down to its children according to these records, so every vertex learns its controller.
    Each member then sends its ID to its controller, starting no earlier than the round after it learns that ID.
    Both passes use $O(1)$ messages per input-edge direction on a tree of height $O(\log n)$ by Lemma~\ref{lem:size-capped-growth}, and each controller receives at most $4B=O(\log n)$ messages.
\end{proof}

\section{Edge responsibilities and virtual-network realization}
\label{app:edge-realization}

\subsection{Proof of Lemma~\ref{lem:mpx-implementation}}
\label{app:mpx-implementation-proof}
\begin{proof}[Proof of Lemma~\ref{lem:mpx-implementation}]
    The endpoints of each input edge first exchange their cluster IDs and their root IDs in $F$ in one round of local communication.
    We execute the MPX procedure of G\"{o}tte et al.~\cite{gotte2023time}, with each controller acting for one cluster vertex.
    In each iteration, the controller distributes the message it sends in MPX to its member vertices, which exchange messages over input edges.
    Each received message has $O(\log n)$ bits, and each member vertex reports only the highest-priority message under the MPX comparison rule, together with the endpoint IDs of the input edge through which it arrived.
    The controller applies the same rule, including any message it generates, and obtains the same result as without filtering.
    When a controller adopts a message proposing another cluster as the center, it records the input edge of that message as the connection to the parent cluster.
    Each iteration takes a constant number of rounds by Lemma~\ref{lem:representative-communication}, so MPX takes $O(\log n)$ rounds in total~\cite[Lemma~4.2]{gotte2023time}.

    Each controller distributes the resulting MPX component ID to its member vertices.
    The member vertices exchange these IDs over input edges.
    For each adjacent MPX component other than its own, each member vertex selects one input edge and reports the opposite component ID, cluster ID, and both endpoint IDs to the controller, one report per round.
    The controller selects one report for each opposite MPX component.
    By Lemma~\ref{lem:mpx-sparsification}, each cluster is adjacent to $O(\log n)$ other MPX components w.h.p., so all reports finish in $O(\log n)$ rounds.
    We impose common bounds of $O(\log n)$ on both the number of reporting rounds and the number of edges selected by each controller.
    Choosing sufficiently large constants and applying a union bound over all clusters ensures that these bounds hold w.h.p.

    For each input edge it selects, the controller notifies the member vertex that reported that edge of its selection.
    Both endpoints of each input edge exchange whether their own cluster selected it.
    We retain the edge only if at least one side selected it and the endpoints have different root IDs in $F$.
    If only one side selected the edge, its cluster is responsible.
    If both sides selected it, the cluster with the smaller ID is responsible.
    Both endpoints therefore agree on responsibility.

    Filtering received messages and edge reports at member vertices bounds each controller's global send count and receive count by $O(B)=O(\log n)$ messages per round.
    Other member vertices also send and receive $O(\log n)$ messages per round, including selection notifications.
    Local capacity $O(1)$ suffices, and global communication uses only known controller and member IDs.
    These operations take $O(\log n)$ rounds in total.
\end{proof}

\subsection{Proofs of Lemmas~\ref{lem:edge-distribution} and~\ref{lem:assigned-edge-communication}}
\label{app:edge-distribution-proof}
\begin{proof}[Proof of Lemma~\ref{lem:edge-distribution}]
    By Lemmas~\ref{lem:mpx-sparsification} and~\ref{lem:cluster-construction}, each cluster assigns $O(\log n)$ representative input edges evenly to $\Theta(\log n)$ member vertices w.h.p., giving $O(1)$ assignments per real vertex.
    Each controller collects the endpoint IDs of only the edges $e=\{u,y\}$ for which its cluster is responsible, from their responsible-side endpoints $u$.
    After determining the assignment, it sends the ordered endpoint IDs $(\ID(u),\ID(y))$ to $p_e$ and the same pair together with the ID of $p_e$ to $u$, which forwards that ID to $y$ over input edge $e$.
    Each cluster handles $O(\log n)$ edges, so global send and receive counts are $O(\log n)$ per real vertex, and the controller and member destination IDs are already known.
    Notifications to $y$ use $O(1)$ messages per input-edge direction, so local capacity $O(1)$ suffices.
    Collection, assignment notification, and notification of the opposite endpoint therefore take a constant number of rounds in sequence.
\end{proof}

\begin{proof}[Proof of Lemma~\ref{lem:assigned-edge-communication}]
    For communication after assignment, $p_e$ and the responsible-side endpoint $u$ use global communication, and $u$ relays messages to and from $y$ over input edge $e$.
    If $p_e=u$, the global step is a local computation.
    Each vertex handles $O(1)$ edges as a responsible vertex and at most the $O(\log n)$ edges assigned to its cluster as a responsible-side endpoint.
    Both its global send count and its receive count are therefore $O(\log n)$.
    Even if $y$ is incident to many representative input edges, relaying uses only a constant number of messages per input-edge direction.
    The assignment notifications introduce all global destination IDs.
    After these notifications finish, traversing the routes in each direction takes $O(1)$ rounds.
\end{proof}

\subsection{Proof of Lemma~\ref{lem:constant-degree-cluster-overlay}}
\label{app:constant-degree-cluster-overlay-proof}
\begin{proof}[Proof of Lemma~\ref{lem:constant-degree-cluster-overlay}]
    Each copy path is attached to $C$ only at $v$, so $T_C$ is a tree.
    The sets $V(T_C)$ partition $V(\widehat H)$, and the \textsf{WFT} transformation preserves each set, proving (1).
    Vertices $p_i$ and $u_i$ lie in the same cluster, hence in the same $F$-component.
    Edge $e_i$ is a representative input edge and therefore joins distinct $F$-components by definition, proving (2).
    By Lemma~\ref{lem:edge-distribution}, each real vertex has $O(1)$ copies and restoration edges.
    Each copy has one restoration edge and $T'_C$ has constant degree, so $\widehat H$ has maximum degree $O(1)$.
    The degree of $v$ in $H$ is at most the sum of the $\widehat H$-degrees of $v$ and its $O(1)$ copies.
    Connectivity of $\widehat H$ follows from (1), (2), and Lemma~\ref{lem:mpx-sparsification}(1), and $H$ is its image under $\sigma$.
\end{proof}

\subsection{Proof of Lemma~\ref{lem:overlay-realization}}
\label{app:overlay-realization-proof}
\begin{proof}[Proof of Lemma~\ref{lem:overlay-realization}]
    Lemmas~\ref{lem:size-capped-growth}, \ref{lem:cluster-construction}, \ref{lem:mpx-implementation}, and~\ref{lem:edge-distribution} provide $F$, $\mathcal P$, and the representative input edges with their responsible vertices in $O(\log n)$ rounds.
    Each responsible vertex $p_i$ creates copy $v_i$ and its restoration-edge record locally.
    The copy ID is an $O(\log n)$-bit tuple consisting of a copy tag, the ID of $p_i$, and the endpoint IDs of representative input edge $e_i=\{u_i,v\}$.
    Vertex $v$ can also compute it.
    Vertex $v$ sends each copy the vertex IDs and simulator IDs of its predecessor and successor on the path, using the input edge for $v\to u_i$ and global communication to a known ID for $u_i\to p_i$.
    Taking each copy's predecessor as its parent, together with the rooting of $F$, gives a rooted tree $T_C$ whose edge endpoints know their connection information and parent--child relation.
    These notifications use $O(1)$ messages per representative input edge, so Lemma~\ref{lem:assigned-edge-communication} establishes the paths in $O(1)$ rounds within the stated capacity.

    Next, we apply Lemma~\ref{lem:tree-to-wft} to each $T_C$ in the network of $O(n)$ vertices including the copies.
    The only virtual edges are path edges, at most two per vertex.
    We therefore take $d=2$ and transform the vertex-disjoint trees $T_C$ in parallel into \textsf{WFT}s $T'_C$ in $O(\log n)$ rounds.
    Each real vertex simulates itself and $O(1)$ copies as distinct vertices, so both the global send count and the receive count increase by only a constant factor and remain $O(\log n)$ per round.
    Communication on $F$ uses local communication along the original input edges, so each input-edge direction still carries $O(1)$ messages.
    Throughout the notifications and transformation, vertex IDs are accompanied by simulator IDs.

    Let $I$ consist of the connection information obtained by the transformation and the restoration-edge records retained at $p_i$.
    Both endpoints of every edge know the vertex and simulator IDs, and $H$ has maximum degree $O(1)$ by Lemma~\ref{lem:constant-degree-cluster-overlay}.
    Thus, mapping neighbors by $\sigma$ realizes $H$ with standard capacity.
\end{proof}

\subsection{Round and capacity bounds for Lemma~\ref{lem:st-restoration}}
\label{app:st-restoration-bounds}
\begin{proof}[Proof of the bounds in Lemma~\ref{lem:st-restoration}]
    Line 1 uses local computation on $I$.
    The tree $\widehat T$ has maximum degree $O(1)$, its endpoint simulators know both endpoint vertex IDs and simulator IDs on every edge, and each real vertex simulates $O(1)$ vertices.
    Thus, line 2 takes $O(\log n)$ rounds by Lemma~\ref{lem:fhs-tree-rooting}.
    Line 3 aggregates on the vertex-disjoint \textsf{WFT}s $T'_C$ in parallel.
    Line 4 sends $O(1)$ messages per selected edge along $p_i\to u_i\to v$ (Lemma~\ref{lem:assigned-edge-communication}).
    Line 5 roots an input-edge tree by Lemma~\ref{lem:fhs-tree-rooting}.
    Every stage uses $O(\log n)$ global sends and receives per real vertex per round and $O(1)$ messages per input-edge direction.
\end{proof}

\section{Few-weight and approximate minimum spanning trees}
\label{app:mst}

\subsection{Round and capacity analysis for Theorem~\ref{thm:kweight-mst}}
\label{app:kweight-analysis}
\begin{proof}[Proof of the bounds in Theorem~\ref{thm:kweight-mst}]
    For one weight, the operations other than the \textsf{ST} calls take $O(\log n)$ rounds.
    They use local capacity $O(1)$ and $O(\log n)$ global sends and receives per real vertex per round.
    Weight sharing needs $\lceil k/p\rceil+1$ exchanges on a constant-degree \textsf{WFT}.
    Each exchange takes $O(\log n)$ rounds with $O(p)$ messages per real vertex per round.
    The $p$ weights of a batch are processed in parallel.
    A batch takes $O(\log n)$ rounds with capacity $(p(\lambda+1),p\gamma)$ as $\gamma=\Omega(\log n)$.
    The total number of rounds is therefore $O((k/p)\log n)$.
    A union bound over the $O(k)\le O(n^2)$ randomized \textsf{ST} calls gives success w.h.p.
\end{proof}

\subsection{Constant-degree graphs}
\label{app:constant-degree-mst}
If the input graph has constant degree, the local communication used in the reduction of Theorem~\ref{thm:kweight-mst} can be replaced by global communication.
\begin{corollary}
    \label{cor:constant-degree-kweight-mst}
    Let $G$ be a realized connected graph of maximum degree $O(1)$, and assume that both endpoints of each edge know its weight.
    If $G$ has at most $k$ distinct weights, then for any $1\le p\le k$, an \textsf{MST} can be constructed with high probability in $O((k/p)\log n)$ rounds in $(0,p\log n)$-\HYBRIDz.
\end{corollary}
\begin{proof}
    If $G$ has no edges, we handle it locally.
    Otherwise, let $k'\ge1$ be the actual number of distinct weights and $p'\coloneqq\min\{p,k'\}$.
    Each connected component of a threshold subgraph is a realized graph of maximum degree $O(1)$.
    We apply Theorem~\ref{thm:known-tree-construction} with a constant bound on the maximum degree.
    The trees of input edges used in the reduction also have constant degree, and their endpoint IDs are known.
    Thus, local communication, including that used for \textsf{WFT} transformation, rooting, and distance computation, can be replaced by global communication.
    In particular, treating these input edges as virtual edges, the all-virtual variant of Remark~\ref{rem:tree-to-wft-variants} with $d=O(1)$ performs the \textsf{WFT} transformation in $(0,\log n)$-\HYBRIDz.
    Processing one weight requires $O(\log n)$ messages for each of sending and receiving per real vertex per round.
    Processing $p'$ distinct weights in parallel therefore fits within capacity $p\log n$.
    Since $k'/p'=\max\{1,k'/p\}\le k/p$, the round bound is $O((k/p)\log n)$, including when $p>k'$.
\end{proof}

\subsection{Proof of Corollary~\ref{cor:approx-mst}}
\label{app:approx-mst-proof}
Edges with the same rounded value $\ell(e)$ belong to the same class.
\begin{proof}[Proof of Corollary~\ref{cor:approx-mst}]
    If $E=\emptyset$, we output the empty tree locally.
    Otherwise, $w(e)\le\ell(e)\le(1+\epsilon)w(e)$, including for weight 0.
    By the rounding argument of Gmyr et al., an \textsf{MST} for the rounded weights is a $(1+\epsilon)$-approximate \textsf{MST} for the original weights.

    Representing each class by its smallest input weight preserves the order of rounded weights using $O(\log n)$-bit values.
    We share weights using the same batch procedure as above.
    Each vertex groups its incident input edges by class and sends the smallest input weight in each group as a candidate.
    We group candidates into classes according to the rounded values computed from their input weights.
    For each class, we aggregate its minimum input weight over the \textsf{WFT} and distribute it as the representative value of the class.

    We apply Corollary~\ref{cor:kweight-mst} to the rounded weights $\ell$ and communicate their representative values.
    For parallelism $p'\le\widehat h_\epsilon$, this uses capacity $(p',p'\log n)$ and $O((\widehat h_\epsilon/p')\log n)$ rounds.
    Since $W=\max(\{2\}\cup\{w(e):e\in E\})$ and $0<\epsilon\le1$, we have
    \[
        \widehat h_\epsilon\le2+\lceil\log_{1+\epsilon}W\rceil
        =O(\epsilon^{-1}\log W).
    \]
    Setting $p'=\min\{p,\widehat h_\epsilon\}$ therefore gives the claimed bound also when $p>\widehat h_\epsilon$.
\end{proof}

\section{Biconnected components}
\label{app:bcc}
Let $T$ be an \textsf{ST} with parent function $p$, call the edges in $E\setminus E(T)$ non-tree edges, let $\pi(v)$ be the preorder number of $v$ with children visited in increasing ID order, and let $D_T(v)$ be the subtree of $v$ with $s_T(v)\coloneqq|D_T(v)|$, so that the preorder numbers of $D_T(v)$ form $[\pi(v),\pi(v)+s_T(v)-1]$.
Define $\operatorname{low}(v)$ and $\operatorname{high}(v)$ as the minimum and maximum of $\{\pi(x):x\in D_T(v)\}\cup\{\pi(y):x\in D_T(v),\{x,y\}\in E\setminus E(T)\}$.
The auxiliary graph $H_{\mathrm{BCC}}$ has one vertex $x_v$ for the parent edge $\{v,p(v)\}$ of each non-root vertex $v$, and an edge $\{x_v,x_w\}$ if and only if (i)~$\{v,w\}$ is a non-tree edge and neither endpoint is an ancestor of the other, or (ii)~$w$ is a child of a non-root vertex $v$ with $\operatorname{low}(w)<\pi(v)$ or $\operatorname{high}(w)\ge\pi(v)+s_T(v)$.

\begin{lemma}[{\cite[Theorem~1 and Steps~3--5]{tarjan1985efficient}}]
    \label{lem:bcc-reduction}
    Assign each tree edge $\{v,p(v)\}$ to the component of $x_v$ in $H_{\mathrm{BCC}}$, and each non-tree edge $\{v,w\}$ with $\pi(v)<\pi(w)$ to the component of $x_w$.
    Two input edges are assigned to the same component if and only if they belong to the same BCC, and every BCC contains a tree edge.
\end{lemma}

\begin{lemma}
    \label{lem:bcc-subtree-values}
    Given an \textsf{ST} in each input component, each vertex $v$ can deterministically compute $\pi(v)$, $s_T(v)$, $\operatorname{low}(v)$, and $\operatorname{high}(v)$ in $O(\log n)$ rounds in $(1,\log n)$-\HYBRIDz.
\end{lemma}

\begin{proof}[Proof of Lemma~\ref{lem:bcc-subtree-values}]
    In a singleton component, we set all four values to 1.
    We assume below that tree $T$ has an edge.
    In an Euler tour, the traversal of each subtree forms one contiguous interval.
    We use this property to compute subtree sizes and extrema by prefix sums and interval aggregation on the sequence.

    Each non-root vertex notifies its parent that the tree edge is selected, and each vertex orders its children by ID.
    We construct a sequence $L$ of directed edges representing an Euler tour that starts and ends at the root.
    Each non-root vertex $v$ is responsible for both $(p(v),v)$ and $(v,p(v))$.
    If a sibling is responsible for a preceding or following item, the parent sends the child that item's ID and its responsible vertex's ID over an input edge.
    Introduced IDs are used from the following round onward.
    We leave the last and first items of $L$ unconnected.
    We use pointer jumping, in which each item introduces its preceding and following reference targets to each other, to obtain the IDs of items at distances $1,2,4,\ldots$ in each direction.
    References outside the sequence are empty.

    Each item from a parent to a child corresponds to the first visit to a non-root vertex.
    We assign 1 to these items and 0 to items from a child to its parent, and let $P$ be the inclusive prefix sum plus 1 to count the root.
    We compute prefix sums by adding sums of disjoint preceding intervals, doubling their lengths through pointer jumping.
    The vertices first visited between the two items of a non-root vertex $v$ are precisely its proper descendants, so
    \[
        \pi(v)=P((p(v),v)),\qquad
        s_T(v)=P((v,p(v)))-P((p(v),v))+1.
    \]
    We set the root's preorder number to 1.
    The root receives the last item's value $P=|V(T)|$ from its responsible vertex over a tree edge and uses it as the component size.

    Next, each vertex exchanges preorder numbers over input edges.
    Each non-root vertex places the minimum and maximum among its own number and those of its non-tree neighbors at both items for which it is responsible.
    The subtree interval of $v$ extends from $(p(v),v)$ to $(v,p(v))$ and contains exactly the two items of each vertex in the subtree.
    Its minimum and maximum are therefore $\operatorname{low}(v)$ and $\operatorname{high}(v)$.
    We use pointer jumping to retain the minimum and maximum at both endpoints of each interval whose endpoints are a power-of-two distance apart.
    Let $m\coloneqq2s_T(v)-1$ be the difference between the positions of the subtree interval's endpoints, and set $j=\lfloor\log m\rfloor$.
    Since $2^j\le m<2^{j+1}$, the two intervals extending $2^j$ hops inward from the endpoints lie within the subtree interval and cover it.
    Overlap does not change the minimum or maximum, so combining their aggregates suffices.
    The subtree of the root $r$ is the entire input component, and no non-tree edge leaves it.
    We therefore set $\operatorname{low}(r)=1$ and $\operatorname{high}(r)=s_T(r)$.

    The vertices responsible for consecutive items in $L$ are the same vertex, adjacent vertices of the tree, or siblings.
    The parent's ID introductions use at most one message per input-edge direction and ensure that the vertices responsible for consecutive items know each other's ID.
    In each pointer-jumping iteration, the ID of the vertex responsible for the new reference target is obtained from the previous reference target and used from the following round onward.
    For each fixed distance and direction, distinct items have distinct nonempty reference targets.
    Since each real vertex is responsible for at most two items, its global send and receive counts are both $O(1)$ per round.
    Tree-edge selection notifications, preorder exchanges, and the return of the component size to the root also use $O(1)$ messages per input-edge direction.
    The sequence $L$ has $2(|V(T)|-1)$ items, so prefix sums and interval aggregation each take $O(\log n)$ rounds.
\end{proof}

\begin{proof}[Proof of Theorem~\ref{thm:biconnectivity}]
    Construct an \textsf{ST} in each input component by Theorem~\ref{thm:low-capacity-st}, compute the four values of Lemma~\ref{lem:bcc-subtree-values}, and exchange them over input edges.
    Following G\"{o}tte et al.~\cite{gotte2023time}, non-root $v$ simulates $x_v$.
    An auxiliary edge of rule (i) is simulated over its non-tree input edge, and one of rule (ii) is simulated over the parent--child input edge.
    Each input edge therefore carries at most one auxiliary edge.
    Both endpoints evaluate the rules consistently.
    The graph $H_{\mathrm{BCC}}$ plus the isolated roots is isomorphic under $x_v\mapsto v$ to a spanning subgraph of the input graph.
    Hence, the \textsf{WFT} construction of Theorem~\ref{thm:low-capacity-st} runs on it with standard capacity.
    Assign each input edge to its endpoint with the larger preorder number.
    For a tree edge, this endpoint is the child, matching Lemma~\ref{lem:bcc-reduction}.
    On the \textsf{WFT} of each component of $H_{\mathrm{BCC}}$, aggregate and broadcast the minimum simulator ID as the BCC label and the number of assigned edges.
    Send both values to the opposite endpoint of each assigned edge.
    A BCC with one edge is a bridge, a vertex with two distinct incident labels is an articulation point, and an isolated vertex is neither.
    Each stage takes $O(\log n)$ rounds, and a union bound over the two randomized tree constructions gives success w.h.p.
\end{proof}
\end{document}